\documentclass[aps,prl,twocolumn,superscriptaddress,fleqn,a4paper]{revtex4-1}
\usepackage{graphicx}
\usepackage[version=3]{mhchem}
\usepackage{chemfig,wrapfig}
\usepackage{float}
\usepackage{siunitx}
\usepackage[hidelinks]{hyperref}
\usepackage{color}
\usepackage{pdfpages}

\makeatletter
\AtBeginDocument{\let\LS@rot\@undefined}
\makeatother

\newcommand{\ybcoF} {$\ce{YBa2Cu3O_{7}}$}
\newcommand{\ybco} {$\ce{YBa2Cu3O_{6+y}}$}
\newcommand{\ybcoU}[1] {$\ce{YBa2Cu3O_{#1}}$}
\newcommand{\ybcoE} {$\ce{YBa2Cu4O8}$}

\newcommand{\tc}[0]{\ensuremath{T_\mathrm{c}}}

\newcommand{\cpara}[0]{\ensuremath{c \parallel B_0}}
\newcommand{\cperp}[0]{\ensuremath{{c\perp B_0}}}
\newcommand{\abpara}[0]{\ensuremath{ab \parallel B_0}}

\newcommand{\degree}[0]{\ensuremath{^{\circ}}}
\newcommand{\nd}[0]{\ensuremath{n_{\mathrm{d}}}}
\newcommand{\np}[0]{\ensuremath{n_{\mathrm{p}}}}

\begin{document}

\title{Proof of bulk charge ordering in the CuO$_2$ plane of the cuprate\\ superconductor YBa$_2$Cu$_3$O$_{6.9}$ by high pressure NMR}

\author{Steven Reichardt}
\affiliation{University of Leipzig, Felix Bloch Institute for Solid State Physics,  
Linnestr. 5, 04103 Leipzig, Germany}
\author{Michael Jurkutat}
\affiliation{University of Leipzig, Felix Bloch Institute for Solid State Physics,  
Linnestr. 5, 04103 Leipzig, Germany}
\author{Robin Guehne}
\affiliation{University of Leipzig, Felix Bloch Institute for Solid State Physics,  
Linnestr. 5, 04103 Leipzig, Germany}
\affiliation{Victoria University of Wellington, The MacDiarmid Institute for Advanced Materials and Nanotechnology, SCPS, PO Box 600, Wellington 6140, New Zealand} 
\author{Jonas Kohlrautz}
\affiliation{University of Leipzig, Felix Bloch Institute for Solid State Physics,  
Linnestr. 5, 04103 Leipzig, Germany}
\author{Andreas Erb}
\affiliation{Walther-Mei\ss ner-Institute for Low Temperature Research, 
Walther-Mei\ss nerstr. 8, 85748 Garching, Germany}
\author{J\"urgen Haase* }
\affiliation{University of Leipzig, Felix Bloch Institute for Solid State Physics,  
Linnestr. 5, 04103 Leipzig, Germany}

\date{\today} 
\begin{abstract}
Cuprate superconductors still hold many open questions, and recently, the role of symmetry breaking electronic charge ordering resurfaced in underdoped cuprates as phenomenon that competes with superconductivity. Here, unambiguous NMR proof is presented for the existence of local charge ordering in nearly optimally doped YBa$_2$Cu$_3$O$_{6.9}$, even up to room temperature. Increasing pressure and decreasing temperature leads to the highest degree of order in the sense that the two oxygen atoms of the unit cell of the CuO$_2$ plane develop a charge difference of about 0.02 holes, and order throughout the whole crystal. At ambient conditions a slightly smaller charge difference and a decreased order is found. Evidence from literature data suggests that this charge ordering is ubiquitous to the CuO$_2$ plane of all cuprates. Thus, the role of charge ordering in the cuprates must be reassessed.

\end{abstract}




\maketitle

 \vspace{0.5cm}

The understanding of the superconducting cuprates is still a pressing problem \cite{Keimer2015}, and it is the charge state of their ubiquitous CuO$_2$ plane that triggers a wealth of electronic properties \cite{Keimer2015,Lee2006,Schrieffer2007}.
Recently, it was discovered that also the sharing of the inherent planar Cu hole with planar O, i.e., the covalency of the bond, can be essential for some properties. For example, the maximal possible critical temperature $\tc$ achieved at optimal doping is nearly proportional to this planar O hole content of its parent (undoped) system \citep{Jurkutat2014, Rybicki2016}. Also recently, charge ordering in the CuO$_2$ plane and its relation to superconductivity or the pseudogap state is again in the focus of research \cite{Comin2014,Campi2015,Blanco-Canosa2014,Ghiringhelli2012,Huecker2014,Chang2012,Wu2011,Wu2013}. 

Nuclear magnetic resonance (NMR) of Cu and O in the CuO$_2$ plane is very sensitive to the local charge symmetry due to the nuclear quadrupole interaction that measures the electric field gradient (EFG) at each nucleus, and one might expect NMR to be a versatile, even benchmark bulk probe for related research. 
However, NMR mostly focused on the electronic spin susceptibility \cite{Schrieffer2007,Walstedt2008,Jurkutat2017,Rybicki2015,Haase2017}, and only scattered publications addressed the charge variation \cite{Hammel1998, Haase2000,Haase2003b,Singer2002}. 
More recent work involves predominantly changes in the local charge symmetry due to rather high magnetic fields for a few selected cuprates \cite{Wu2011,Wu2013,Wu2015,Kharkov2016,Reichardt2016}. 
The chief reason for why NMR has not been dominating in this field is perhaps the following. Two cuprates, \ybcoF{} and \ybcoE{}, appear to be very 
\begin{figure}[h!]
 \centering
 \includegraphics[scale=0.56]{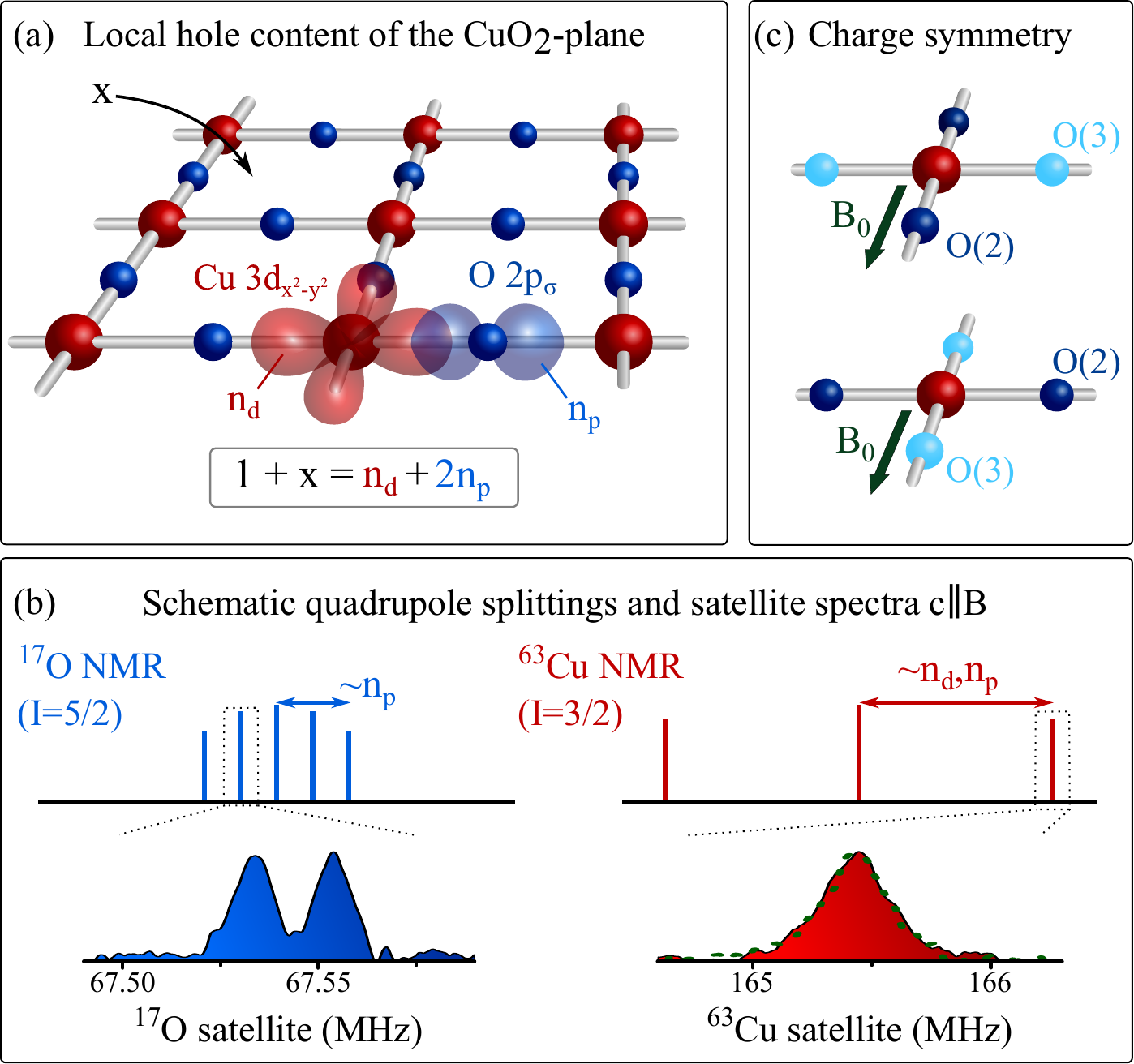}
 \caption{NMR and charge distribution in the CuO$_2$ plane. (a) $^{63,65}$Cu and $^{17}$O NMR can measure the hole contents of the bonding orbitals, $\nd$ (3d$_{x^2-y^2}$) and $\np$ (2p$_{\sigma}$) through the quadrupole splitting, sketched in the upper part of (b), which shows that the oxygen splitting is proportional to $\np$, while that of Cu depends linearly on $\np$ and $\nd$; NMR recovers the total hole content $1+x$ (inherent plus doped (x) holes). Typical O and Cu satellite spectra of the investigated sample YBa$_2$Cu$_3$O$_{6.9}$ at ambient conditions for \cpara{} are shown in the lower part of (b). While the Cu spectrum has a lineshape that agrees with the measured line in zero field (green dashed), the oxygen satellite is split indicating different planar O sites (O(2,3)).  (c) With an in-plane magnetic field we can distinguish the spatial arrangement of O(2,3) and we find charge ordering to be the origin, rather than the crystal's orthorhombicity.}
 \label{fig:Fig1}
\end{figure}
homogeneous, in the sense that the measured components of the EFGs hardly vary across the CuO$_2$ plane, i.e., one observes very narrow NMR lines for planar Cu and O in these stoichiometric systems. 
Basically all the other cuprates show broad, featureless Cu and O lines, whose widths can vary with sample preparation or doping \cite{Reichardt2016}. 
With a few materials being very homogeneous, the broad, featureless lines of all the other systems were usually dismissed as being related to impurities or crystal imperfections, poorly shielded. 
 
The above view of NMR, however, rests on an important assumption. \ybcoF{} and \ybcoE{} show, e.g., a substantial line splitting for planar O, cf.~Fig.~\ref{fig:Fig1}~(b), which is attributed to the orthorhombicity of these materials as it makes planar O in the $a$ and $b$ directions different from each other. While reasonable, it was pointed out years ago that if this interpretation fails, these splittings could easily signal large charge density variations of particular symmetry \cite{Haase2003,Haase2003b}.

Here, we tackle this long-standing issue with new high-pressure NMR experiments \cite{Haase2009,Meissner2011}. We use pressure high enough to affect the electronic properties of the material, but low enough to leave the chemical structure and crystal twinning unchanged. This lets us separate electronic spatial variations from those due to the chemical structure. 
Our results show unambiguously that electronic charge ordering at planar oxygen is present, which reacts to pressure and temperature. It is known that various members of the \ybco{} family of materials (YBCO) show the presumed 'orthorhombic' planar O splitting to various degrees. 
Thus, the reported charge ordering appears to be ubiquitous to this family of materials. 
Furthermore, we present literature data on a number of cuprates, which suggest that this charge ordering is ubiquitous to the CuO$_2$ plane of all cuprates. 

\section{Results}
\subsection{Charge ordering and NMR}
Before we present the results for this important proof, we explain how NMR measures the charge distribution in the CuO$_2$ plane. 
$^{63,65}$Cu or $^{17}$O nuclear spins interact due to their quadrupole moments with electric field gradients (EFG) present at the nuclear sites. With the moments known, NMR can measure the principle components ($V_{XX},V_{YY},V_{ZZ}$) of the EFG tensors, as well as their orientations in the crystal (at Cu and O sites), cf. Suppl.~Fig.~1 and 2. 

For Cu with the partly filled $3d_{x^2-y^2}$ orbital the EFG tensor must be nearly symmetric with its largest component ($V_{ZZ}$) parallel to the crystal $c$-axis, and $V_{XX} \approx V_{YY} \approx -V_{ZZ}/2$ (traceless tensor), with $X$ and $Y$ along the planar $a$ or $b$ axes.
Since the quadrupole interaction is quite strong for Cu, we used nuclear quadrupole resonance (NQR) to determine $V_{ZZ}$ without a magnetic field, as well.
For planar O the situation is different, cf.~Fig.~\ref{fig:Fig1}. 
A full shell configuration, O$^{2-}$, produces a vanishing EFG, but the hybridization with Cu results in some hole content in the $2p_\sigma$ orbital so that the largest principle value for the O EFG (while small compared to Cu) will be along this $\sigma$-bond, i.e., along Cu-O-Cu bonds. 
The out of plane chemistry will make the O EFG asymmetric.
For the planar CuO$_2$ unit cell the two O sites have their Cu-O-Cu bonds at right angles to each other. 
With the magnetic field along one of the Cu-O-Cu bond directions, cf. Fig.~\ref{fig:Fig1}~(c), the O site with Cu-O-Cu perpendicular to it resonates in a rather different spectral range, and the signals can easily be separated form each other, cf. Suppl.~Fig.~2.

It is important to point out that we must use \textit{twinned single crystals} in order to have sufficiently high precision for comparing spectra from the crystal $a$ and $b$ axes, without systematic error from crystal alignment procedures.
Note that for any chosen direction of the magnetic field in the CuO$_2$ plane the axes $a$ and $b$ have projections with equal likelihood on the field direction. 
Thus, if the orthorhombicity were the determining factor for the observed NMR line splittings, both directions must contribute equally to every NMR spectrum, cf. Suppl.~Fig.~1. We provide sufficient proof in Methods for why we can exclude changes in the chemical structure including de- or re-twinning during the course of our experiments.

We will report below that as a function of pressure and temperature the EFGs at Cu and O order (throughout the whole crystal). This proves the ordering of the charges in the CuO$_2$ plane that we can quantify due to very recent progress relating the averge EFGs in the plane to the hole contents of the bonding orbitals, $\nd$ of Cu 3$d_{x^2-y^2}$, and $\np$ of O $2p_\sigma$. NMR can measure the actual doping ($x$) quite reliably from determining the EFGs, and the expected relation, $1+x = \nd +2 \np$ for hole ($x>0$) as well as for electron ($x<0$) doped systems is recovered, i.e., $\nd$ and $\np$ add up to the sum of inherent Cu hole plus doped hole ($1+x$)  \cite{Jurkutat2014,Rybicki2016}, cf. also Fig.~\ref{fig:Fig1}. We find with this simple formalism that the ordering of the EFGs at Cu and O is explained quantitatively by charge ordering of the 2p$_\sigma$ orbitals.

\subsection{Ordering Phenomena from $^{63}$Cu NMR}
\begin{figure}[t]
 \centering
 \includegraphics[scale=0.6]{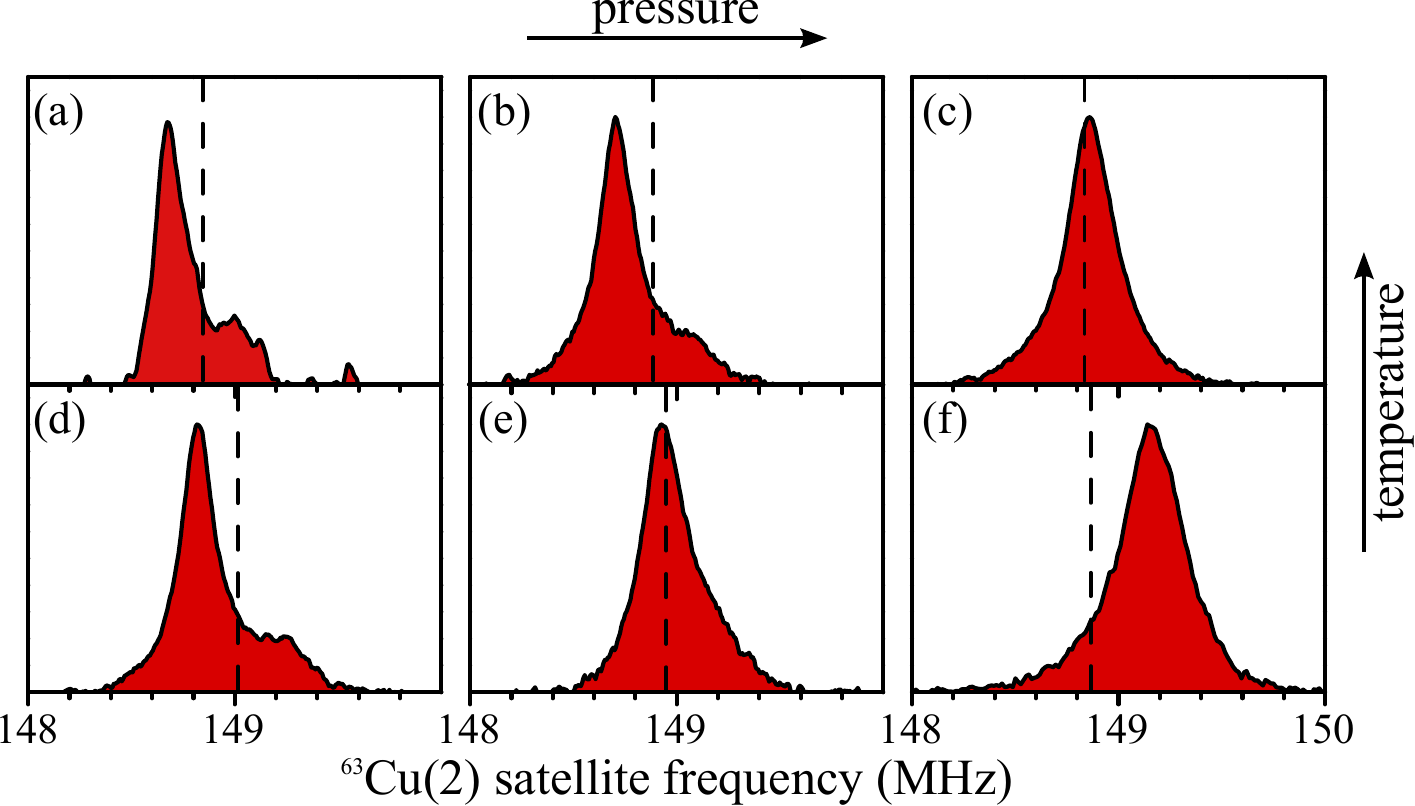}
\caption{Planar $^{63}$Cu satellite spectra for \abpara{} at various pressures: ambient (a, d); \SI{9}{kbar} (b,e); \SI{18}{kbar} (c,f); at \SI{300}{K} and \SI{100}{K} for the upper and lower panel, respectively. Vertical dashed lines show the calculated position for an axially symmetric EFG (with zero field data). Panels (a, d, b): most nuclei resonate at $\Delta f_- = - \epsilon$; panel (c, e) $\Delta f \approx 0$; panel (f) $\Delta f_+ \approx +2 \epsilon$ ($\epsilon \propto |V_{XX}-V_{YY}|$). These observations reveal a change in the asymmetry of the Cu EFG with pressure and temperature.}
 \label{fig:Fig2}
\end{figure}
For the planar Cu NMR satellites, cf. Fig.~\ref{fig:Fig1}~(b), the magnetic contributions to the linewidths are negligible, which makes data analysis very reliable, and we present our results sorted by how the magnetic field was aligned with respect to the crystal axes. 

{\flushleft \textit{(1) Magnetic field $B_0$ parallel to crystal $c$-axis (\cpara{}):}}
For planar $^{63}$Cu we measure $V_{ZZ}$ and its spatial distribution for this orientation.
The results are compared with NQR in Fig.~\ref{fig:Fig1}~(b), and we find nearly identical lineshapes at ambient conditions (\SI{0.3}{MHz} width). We deduce a quadrupole frequency (proportional to ${V_{ZZ}}$) of $\SI{31.17}{MHz}$, consistent with literature data \cite{Mali1987}. 
Upon applying pressure, $V_{ZZ}$ changes only slightly ($\approx 1$ \%), while its spatial variation (linewidth) increases by a factor of two at \SI{18}{kbar} (Suppl.~Fig.~3). 
In summary, apart from a slight increase in broadening, pressure and temperature have only small effects on the principle EFG component $V_{ZZ}$.

{\flushleft \textit{(2) Magnetic field $B_0$ in the plane (\abpara{}):}\;\;}
We now turn the sample by 90$^{\circ}$ such that the magnetic field is along $a$ and $b$ for this  twinned crystal (Cu-O-Cu bond alignment set with $^{17}$O NMR, see Methods). Selected spectra of the high frequency Cu satellite are shown in Fig.~\ref{fig:Fig2}. 
The dashed vertical lines in Fig.~\ref{fig:Fig2} denote the calculated resonance line positions for an exactly symmetric EFG, i.e.,  $V_{XX} = V_{YY}$  (using $V_{ZZ}$ from NQR/NMR). A Cu nucleus with a slightly asymmetric tensor would resonate at a certain frequency distance above ($\Delta f_+=+\epsilon$) or below ($\Delta f_-=-\epsilon$) that line, depending on the orientation of the EFG with respect to the in-plane magnetic field ($\epsilon$ is proportional to $|V_{XX}-V_{YY}|$), cf. Suppl.~Fig.~2. 
If a slightly asymmetric EFG were fixed by the orthorhombic distortion, we would see two resonance lines of similar intensity, namely one at $\Delta f_+$ and the other at $\Delta f_-$. 
This is not observed in Fig.~\ref{fig:Fig2}.

At ambient conditions, cf. Fig.~\ref{fig:Fig2} (a), we find most spectral weight below the dashed line at $\Delta f_-$. 
The intensity at $\Delta f_+$ is too small to be caused by twinning (see also Suppl.~Fig.~4).
At $\SI{9}{kbar}$ and room temperature, cf. Fig.~\ref{fig:Fig2}~(b), the lineshape remains similar. 
At $\SI{18}{kbar}$ and 300$\,$K, the lineshape is nearly symmetric and the center of gravity is close to the dashed line, cf. Fig.~\ref{fig:Fig2}~(c), similar to the situation found already at \SI{9}{kbar} and \SI{100}{K}.
The linewidths are almost precisely what is expected from NQR (cf. Suppl.~Fig.~4).
Thus, all Cu sites have a nearly symmetric tensor  ($\Delta f \approx 0$). 
At $\SI{18}{kbar}$ and $\SI{100}{K}$, cf. Fig.~\ref{fig:Fig2}~(f), the resonance line is shifted to a higher frequency ($\Delta f_+$), saying that basically all Cu sites have a similar asymmetry, almost twice that found at ambient conditions, but of different sign. 
This means, all EFGs are now ordered throughout the whole crystal, not only in $c$-direction but also in the plane irrespective of orthorhombicity.

\begin{figure}
 \centering
 \includegraphics[scale=0.7]{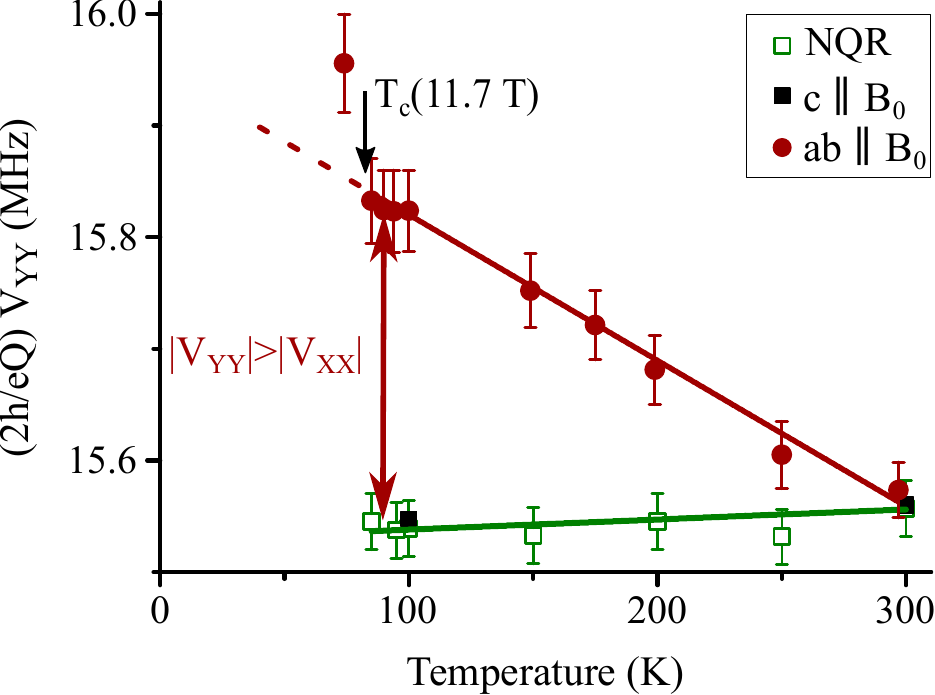}
\caption{Temperature dependence of the planar Cu EFG component $V_{YY}$ (red) measured  at $18\,$kbar; also shown is  $V_{YY}=-V_{ZZ}/2$ determined by NQR (green), and from NMR in \cpara{} (black) as expected for an axially symmetric Cu EFG. This shows a (reversible) linear change of the Cu EFG symmetry as a function of temperature between 300$\,$K and $100\,$K. Below \tc{} only one point could be measured due to screening/sensitivity issues.}
 \label{fig:Fig3}
\end{figure}

Note that a change in orthorhombicity would require the Cu-O chains to change their orientation, which is of course not the case for such low pressures and temperatures (in particular the temperature dependence of the Cu EFG asymmetry in Fig.~\ref{fig:Fig3} would not be reversible, see also Suppl.~Fig.~5).
More importantly, we can exclude changes due to misalignment since this would shift the resonances to lower quadrupole frequencies. We also tracked the NMR intensities, and there is no intensity loss observed between \SI{0}{kbar} and \SI{18}{kbar}.

Thus, the data represent a simple charge ordering phenomenon. At \SI{18}{kbar} and \SI{100}{K} the whole crystal has the same local order. As the temperature is raised, the average tensor becomes symmetric, but at ambient pressure we observe again charge ordering, predominantly of opposite sign and almost half the amplitude.

\subsection{Ordering Phenomena from $^{17}$O NMR}
\begin{figure}
 \centering
 \includegraphics[scale=0.55]{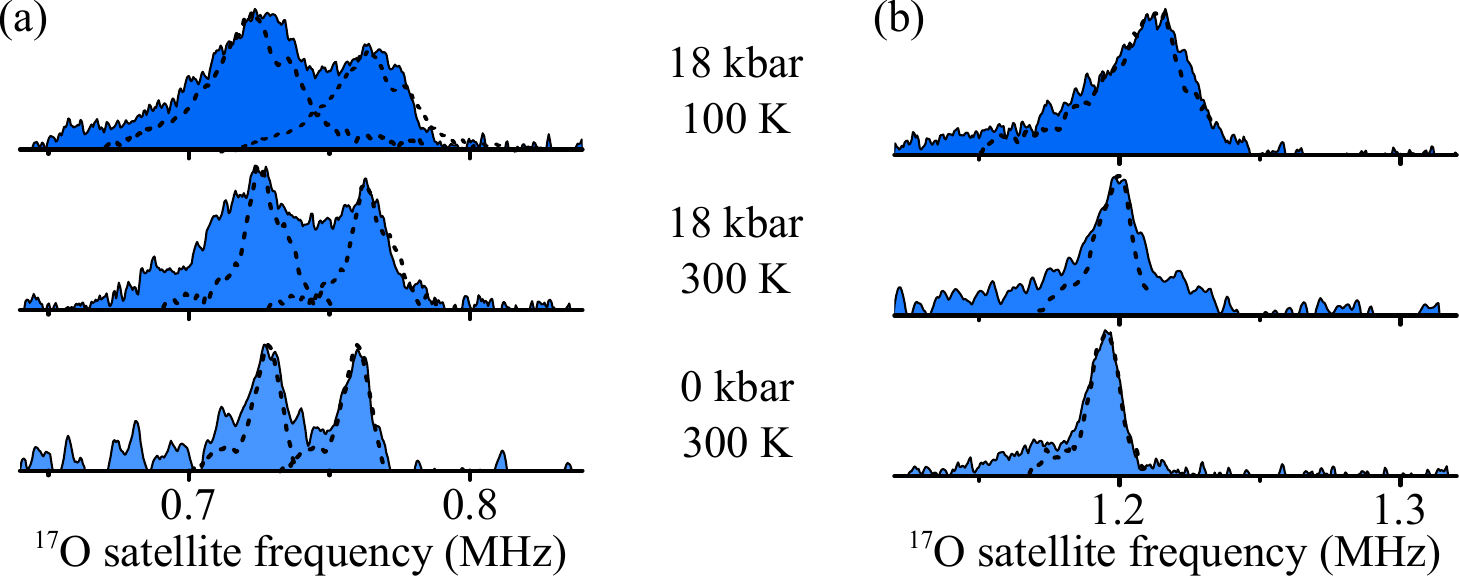}
 \caption{Pressure and temperature dependence of the $^{17}$O uppermost satellite transition (magnetic shifts have been subtracted); (a) magnetic field along the crystal $c$-axis; (b) magnetic field within the plane and perpendicular to the planar O $\sigma$ bonds (Cu-O-Cu). The respective central transitions (shifted in frequency) that only experience magnetic broadening are plotted for each peak as well; note, while magnetic variations contribute one-time, the EFG variations contribute twice to the linewidth of the uppermost O satellite.}
 \label{fig:Fig4}
\end{figure}
For $^{17}$O NMR, the quadrupole splittings are correlated with magnetic shifts, and at magnetic fields typical for NMR both are similar in size. This lead to asymmetric $^{17}$O lineshapes in the cuprates \cite{Haase2000}, and introduces some uncertainty in the data analysis. We determined the approximate average magnetic shifts and subtracted them in the spectra presented, but the linewidths are unaltered. As discussed above, for planar O its $V_{ZZ}$ is found along the Cu-O-Cu bonds, and $V_{XX}$ along the crystal $c$-axis, and $(V_{XX}-V_{YY})/V_{ZZ}\approx 0.23$ \cite{Takigawa1989}.

{\flushleft \textit{(1) Magnetic field $B_0$ parallel to crystal $c$-axis (\cpara{}):}}
The satellite lines for \cpara{} are split, i.e., there are two distinct quadrupole frequencies, cf.~Fig.~\ref{fig:Fig1}~(b). 
This shows that there are two different oxygen sites with similar abundance. 
More data for this orientation at different pressures and temperatures are shown in Fig.~\ref{fig:Fig4}~(a).
The peak positions do not change significantly with increasing pressure or decreasing temperature, but we observe an overall broadening that is largely magnetic at \SI{100}{K}. 
The peak splittings corresponds to quadrupole frequency differences of \SI{18}{kHz} to \SI{26}{kHz}.

Qualitatively, this is in agreement with what one expects from two inequivalent planar O sites due to orthorhombicity, the explanation adopted early on \cite{Takigawa1989, Horvatic1989}.

{\flushleft \textit{(2) Magnetic field $B_0$ in the plane (\abpara{}):}\;\;}
Note, that this orientation is identical to that used for Cu NMR, only the radio frequency tuning outside the anvil cell was changed. The signals from planar O with the $\sigma$-bond parallel ($\sigma \parallel B_0$) and perpendicular ($\sigma \perp B_0$) to the field ($B_0$) can easily be separated ($V_{ZZ}$ and $V_{YY}$ are rather different) and we discuss them separately. 
Unfortunately, we were not able to rotate the field within the plane due to experimental constraints.

\textit{Oxygens with $\sigma$ bonds perpendicular to $B_0$ ($\sigma \perp B_0$)}:
Examples of spectra are shown in Fig.~\ref{fig:Fig4}~(b).
Note that the clear double peak structure as for \cpara{} in Fig.~\ref{fig:Fig4}~(a) is absent, and the line becomes rather broad at higher pressure and lower temperature.
The linewidth at \SI{18}{kbar}, \SI{100}{K} is dominated by magnetic broadening (black dashed line). In our twinned sample, we would expect two lines with similar intensities since $a$ and $b$ axes are equally likely to be perpendicular to the field, cf. Suppl.~Fig.~1. Therefore, the spectra argue against orthorhombic distortion as the origin for the split \cpara{} O lines.

\begin{figure*}[t]
 \centering
 \includegraphics[scale=1]{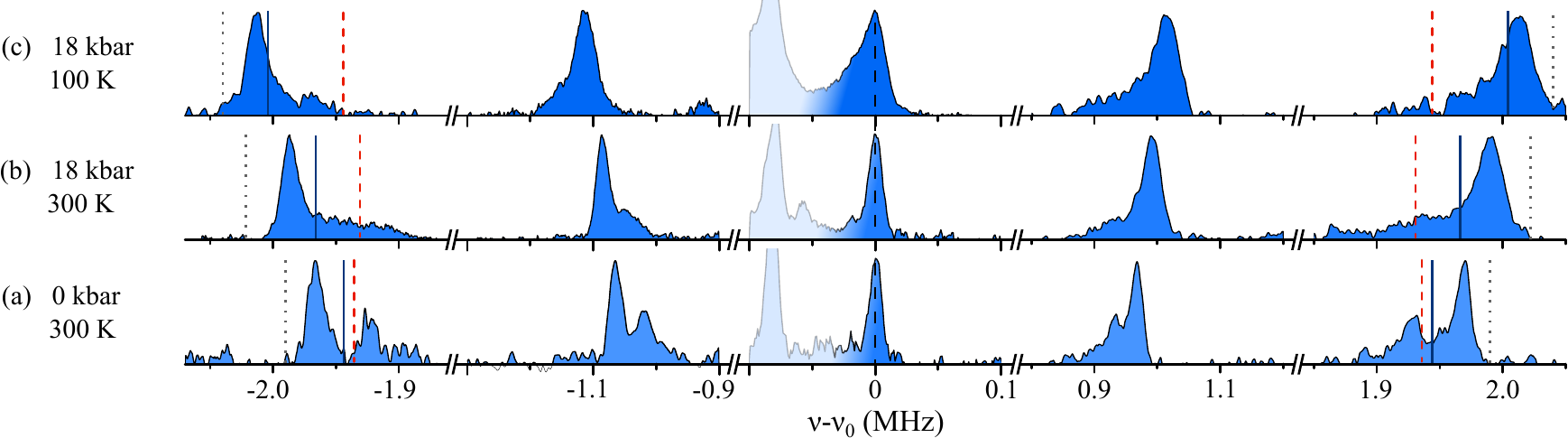}
   \caption{Pressure and temperature dependence of the five planar $^{17}$O NMR resonances for O atoms that have the magnetic field parallel to their $\sigma$ bonds (parallel Cu-O-Cu). Unrelated resonances are shaded. The magnetic shift is subtracted. The blue solid lines are centers of gravity of the outermost transitions (for which the EFG distribution is most pronounced). The red dashed and grey dotted lines were calculated from the outermost satellites of the other two field orientations, e.g., shown in Fig.~\ref{fig:Fig4}. While the grey dotted lines show the consistency of the spectra, the calculated and actual centers of gravity are very different at higher pressure, as expected from local charge ordering (see text and methods for details).}
  \label{fig:Fig5}
\end{figure*}

\textit{Oxygens with $\sigma$ bonds parallel to $B_0$ ($\sigma \parallel B_0$)}: For these O atoms pronounced changes are observed, so we show in Fig.~\ref{fig:Fig5} all five $^{17}$O transitions for clarity.
At ambient conditions, Fig.~\ref{fig:Fig5} (a), there is a double-peak structure. 
The number of nuclei in each environment could be similar if one considers the different linewidths, cf. Suppl.~Fig.~6.
The line splitting is close to what we measured for \cpara{}, and therefore only about half of what is expected for $V_{ZZ}$ along $B_0$. But the total spectral range of a satellite line is indeed about twice as large as for the other two orientations, cf. Fig.~\ref{fig:Fig4}.

At $\SI{18}{kbar}$, Fig.~\ref{fig:Fig5} (b), the line with the smaller quadrupole frequency broadens substantially while the other line remains similar in shape.
The overall quadrupole frequency increases.
From the linewidth and relative intensities we deduce that about half of the nuclei experience a well defined EFG compared to the other half that show a smaller EFG with large spatial variation.
As we lower the temperature at \SI{18}{kbar} to \SI{100}{K}, Fig.~\ref{fig:Fig5}~(c), the broad peak nearly merges with the well defined peak that has now considerably more intensity ($>$ 75\% of the total intensity) and shows an even larger quadrupole frequency. 

This shows that planar O atoms that have their two $\sigma$-bonds parallel to the field undergo pressure and temperature dependent ordering of the planar EFGs, which is most pronounced at \SI{18}{kbar} and \SI{100}{K}.
Such spectral changes, that are reversible and have one peak altered while the other stays rather similar, clearly rule out a structural origin, and thus, must be due to electronic ordering.

\begin{figure}
 \centering
 \includegraphics[scale=0.7]{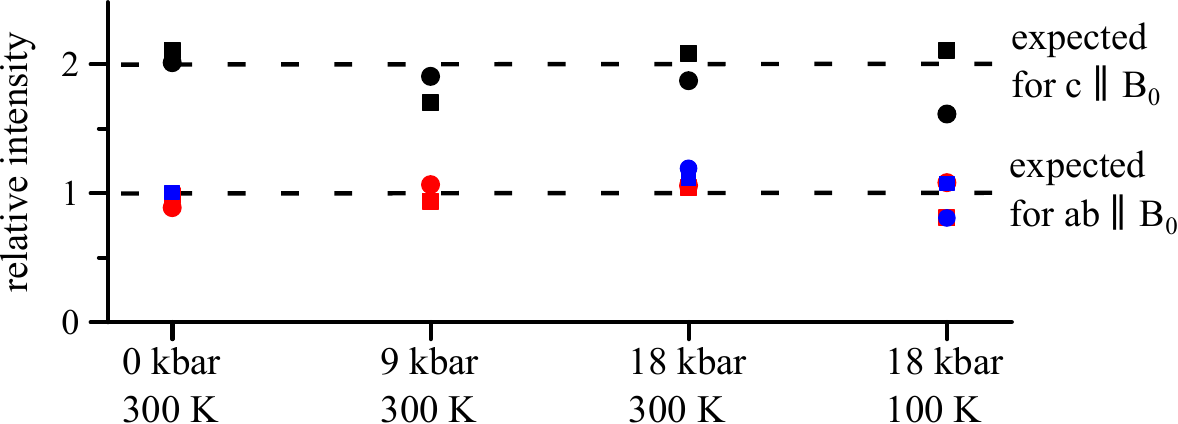}
\caption{Measured total $^{17}$O NMR intensity for the uppermost (square) and lowermost (circle) satellite transitions for the three different orientations of the oxygen $\sigma$ bonds (Cu-O-Cu) with respect to the magnetic field: $\sigma \parallel B_0$ (red), $\sigma \perp B_0$ (blue), and \cpara{} (black). Note that for \cpara{}, both planar O sites are measured simultaneously and one expects the intensity for this direction to be the sum of the intensities for the other two directions.}
 \label{fig:Fig6}
\end{figure}

{\flushleft \textit{(3) Intensity and spectral distribution}:\;\;}
In order to be able to draw firm conclusions we determined the $^{17}$O NMR intensities for all orientations, see Fig.~\ref{fig:Fig6}. As exepected, the number of nuclei contributing to the spectra for \cpara{} is about twice as large as that for either set of spectra for \abpara{}. Thus, within small error bars we can be certain that \textit{all} spectra represent true histograms of the local EFGs at planar O in the CuO$_2$ plane. 

We now calculate two characteristic frequencies, cf. Methods. 
First, we determine the high frequency cut-off from the data in Fig.~\ref{fig:Fig4}, and the gray dotted lines in Fig.~\ref{fig:Fig5} denote the resulting positions for this alignment. The agreement is good and shows the consistency of the data. Second, we calculate the centers of gravity. 
From the data in Fig.~\ref{fig:Fig4} (including transitions not shown) we calculate centers of gravity denoted by red dashed lines in Fig.~\ref{fig:Fig5}. We compare these with the actual centers of gravity from the spectra in Fig.~\ref{fig:Fig5}, the full blue lines. Clearly, there is stark contrast at higher pressure, substantial already at \SI{18}{kbar}, \SI{300}{K}, but at \SI{18}{kbar} and \SI{100}{K} the expected frequencies are even below the actual low frequency cut-off (when considering peak positions, we find similar behavior, cf. Suppl.~Fig.~7).
These discrepancies prove that spectral weights measured for the three different directions of the field cannot come from the same nuclear sites.
Furthermore, since all nuclei contribute to a given \cpara{} satellite, cf. Fig.~\ref{fig:Fig6}, it must be that the two signals for the in-plane resonances (parallel and perpendicular to the field) do not belong to the same O atoms at \SI{18}{kbar} and \SI{100}{K}. In other words,  planar O atoms parallel to the field are different from those perpendicular to the field. Note, again, this order is observed in a twinned crystal, thus, this ordering of the O EFGs cannot be attributed to orthorhombicity.

\subsection{Charge Ordering in the CuO$_2$ Plane}

\begin{figure}
 \centering
 \includegraphics[scale=0.23]{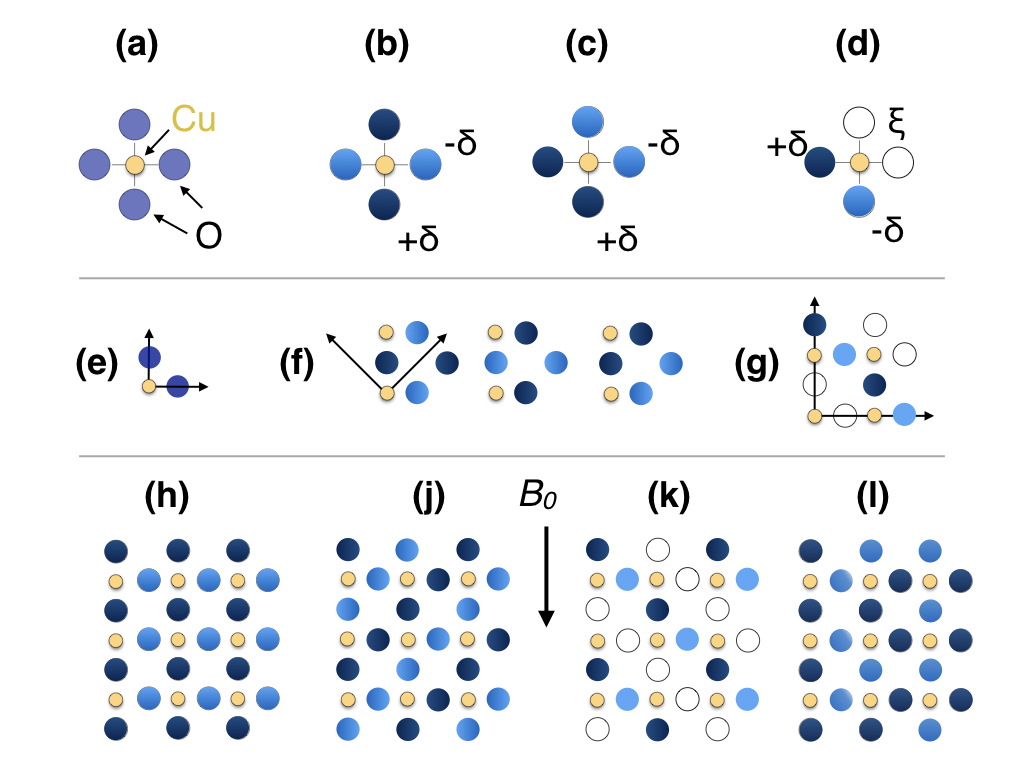}
\caption{Charge ordering scenarios. (a) a CuO$_4$ unit for which in (b,c,d) excess charges $\pm \delta$ are arranged to give (b) a finite Cu asymmetry; (c) a vanishing Cu asymmetry, and (d) a possible intermediate distribution that has a not well-defined charge $\xi$ ($|\xi|<\delta$), in addition to those with $\pm \delta$. The unit cell of the homogeneous CuO$_2$ plane (e) is doubled (Cu$_2$O$_4$) in (f) for three fundamentally different charge ordering scenarios; (g) shows a Cu$_4$O$_8$ unit cell. Corresponding long-range patterns are shown in (h) through (l). For the direction of the magnetic field $B_0$, (h) is in agreement with the data at \SI{18}{kbar}, \SI{100}{K}; (j) results in a vanishing Cu EFG, but the two O positions for \abpara{} are not observed, and (k) gives a possible solution as it results for \abpara{} in two distinct O signals from ordered charges and an incommensurate distribution of charges $\xi$. Panel (l) would be in agreement with the measurements at ambient conditions.}
 \label{fig:Fig7}
\end{figure}

Since the average Cu and O EFGs are determined by the average charge in the bonding orbitals of the CuO$_2$ plane \cite{Haase2004,Jurkutat2014}, cf. Fig.~\ref{fig:Fig1}~(a), we will attempt to describe the spatial charge ordering quantitatively by this formalism, as well.
For this, we must consider CuO$_4$ units, cf. Fig.~\ref{fig:Fig7}~(a), with a Cu hole content $\nd$ and O hole contents $n_{\mathrm{p},1...4}$, since for the Cu EFG also the charges at the four neighboring O atoms contribute \cite{Haase2004}. 

\textit{(1) \SI{18}{kbar}, \SI{100}{K}}: With the magnetic field in the plane we find the EFG tensors at all Cu atoms aligned in the plane, i.e., with asymmetry $\Delta f_+$ (and their $V_{ZZ}$ along the $c$-direction). Under exactly the same conditions the O EFGs are aligned, as well. Here, the atoms with their $\sigma$-bonds parallel to the field experience a larger field gradient. Thus, these $\sigma$-orbitals must carry a slightly higher charge \cite{Haase2004}. We define it to be $+\delta$, and denote with $-\delta$ the smaller charge at the O atoms that have their $\sigma$-bonds perpendicular to the in-plane magnetic field.
This arrangement is depicted in Fig.~\ref{fig:Fig7}~(b), and amounts to a charge modulation within the unit cell of the homogeneous CuO$_2$ plane. With two different O charges we can double the unit cell and have a choice of arranging the $\pm \delta$ charges, three fundamentally different possibilities are shown in Fig.~\ref{fig:Fig7}~(f). The first unit cell under (f) will result in a charge density wave (h) consistent with our data.
From the O data we determine \cite{Haase2004} an amplitude of the charge density variation in the O $2p_\sigma$ orbital of about $\delta = 0.01$ (holes per O).
This amplitude with the configuration given in Fig.~\ref{fig:Fig7}~(h) results in a Cu EFG asymmetry with a positive frequency shift of \SI{+340}{kHz} (see Methods for details), in  agreement with the experimentally found asymmetry shift of \SI{+320}{kHz} in Fig.~\ref{fig:Fig2}~(f). 
Note that while we can be certain about bulk ordering, if less than about 15\% of the Cu atom were missing in our spectra (due to excessive broadening) we would find the same result since such a loss is almost unnoticed in these difficult NMR experiments. Thus, on average the ordering could be locally quite incommensurate, but globally it is quite  commensurate. Of course, different parts of the sample could have different wave vectors, e.g., due to impurities. 

\textit{(2) \SI{18}{kbar}, \SI{300}{K}}: As the temperature is increased, Fig.~\ref{fig:Fig3}, the asymmetry decreases linearly, and at \SI{300}{K} the spectrum is nearly symmetric. A symmetric Cu EFG results if charges at the four O neighbors are the same, however, this is also the case for a distribution according to Fig.~\ref{fig:Fig7}~(c). Note that the two O lines for \cpara{} demand two O atoms of similar abundance with slightly different charge. The third unit cell in Fig.~\ref{fig:Fig7}~(f) would result in a charge density wave according to (j). However, this would demand two peaks for the O spectra in \abpara{}, which we do not find. For O with $\sigma$-bonds parallel and perpendicular to the field we find one well defined line each, for which the differences in the centers of gravity still prove order. The second broad spectral feature hints at disorder, but the centers of gravities are similar. In order to understand this as a long-range pattern, a larger unit cell is necessary. One such cell is indicated in Fig.~\ref{fig:Fig7}~(g). The two well-defined sites (with $\pm \delta$) carry the obvious colors, while the disordered sites (($|\xi|<\delta$)) are left empty. A possible pattern consistent with our data is presented in Fig.~\ref{fig:Fig7}~(k).

\textit{(3) \SI{9}{kbar}}: At $100\,$K, we face perhaps a situation similar to that at \SI{18}{kbar}, \SI{300}{K} (a complete set of O data is not available), except that the average Cu asymmetry is even closer to zero, or has begun to change sign. The discussion would be similar to the one outlined above. At \textit{\SI{9}{kbar}, \SI{300}{K}} the Cu asymmetry has changed sign and remains similar at lower pressure at temperatures above \SI{100}{K}. Therefore, we discuss these lineshapes in the next paragraph.

\textit{(4) \rm Ambient pressure}: Similar to \SI{9}{kbar}, \SI{300}{K} a planar EFG asymmetry at Cu is found, but with a different sign for most of the nuclei, and is about half in size compared to \SI{18}{kbar}, \SI{100}{K}. Only a smaller fraction of nuclei shows the high pressure, low temperature asymmetry. The overall linewidth has decreased slightly.
A combination of the first two unit cells in Fig.~\ref{fig:Fig7}~(f) can describe the asymmetry sufficiently well, and the charge difference between the O atoms is about 70\% of that measured at \SI{18}{kbar}, \SI{100}{K}. The O spectra in Fig.~\ref{fig:Fig5} show a two-peak structure, as for \cpara{}, but for the O atoms with $\sigma$-bonds perpendicular to the in-plane field, Fig.~\ref{fig:Fig4}~(b), the second line, if present, appears to be smeared over a broader spectral range. A simple scenario is sketched in Fig.~\ref{fig:Fig7}~(l) (where we used the same colors while $\delta$ has decreased). Likewise, parts of the sample could order with either unit cell so that locally a higher order could be present. 
Besides the orientation of the Cu asymmetry, its amplitude matches the respective O satellite line splittings throughout, e.g., at ambient conditions we deduce an O charge variation amplitude of $\delta \approx 0.007 $ that, if ordered, would result in a deviation $\Delta f_{\pm}$= \SI{\pm250}{kHz}, in agreement with the experimentally measured deviation shown in Fig.~\ref{fig:Fig2}~(a,b,d).

\section{Discussion}
The driving force for this comprehensive set of experiments was to find out whether those cuprates with rather narrow NMR lines are indeed very homogeneous, or if, e.g., the split $^{17}$O NMR satellites signal special intra unit cell charge ordering, as proposed some years ago \cite{Haase2003,Haase2003b}.
Such a charge density variation that breaks the symmetry of the chemical lattice may show a special behavior under pressure, and indeed, our results on YBa$_2$Cu$_3$O$_{6.9}$ reveal such a special electronic response, i.e., the electric field gradient tensors of Cu and O in the plane undergo different ordering scenarios as function of pressure and temperature. With the recently established quantitative picture for the measurement of the average charges with NMR \cite{Haase2004, Jurkutat2014} we find that a charge modulation of the O $2p_\sigma$ orbitals of $\lesssim$ 1\% is behind the observed EFG modulations. The ordering can be influenced already at $9\,$kbar by moderate temperature changes (note that \SI{9}{kbar} correspond to an energy density of a \SI{\sim 50}{Tesla} field, at which about 1\% of the total O hole content is being spatially rearranged, not far from \tc{}). With the bulk ordering observed, chemical inhomogeneity cannot be responsible for the effects, while it may be involved in pinning. 

\begin{figure}[t]
 \centering
 \includegraphics[width=0.45\textwidth]{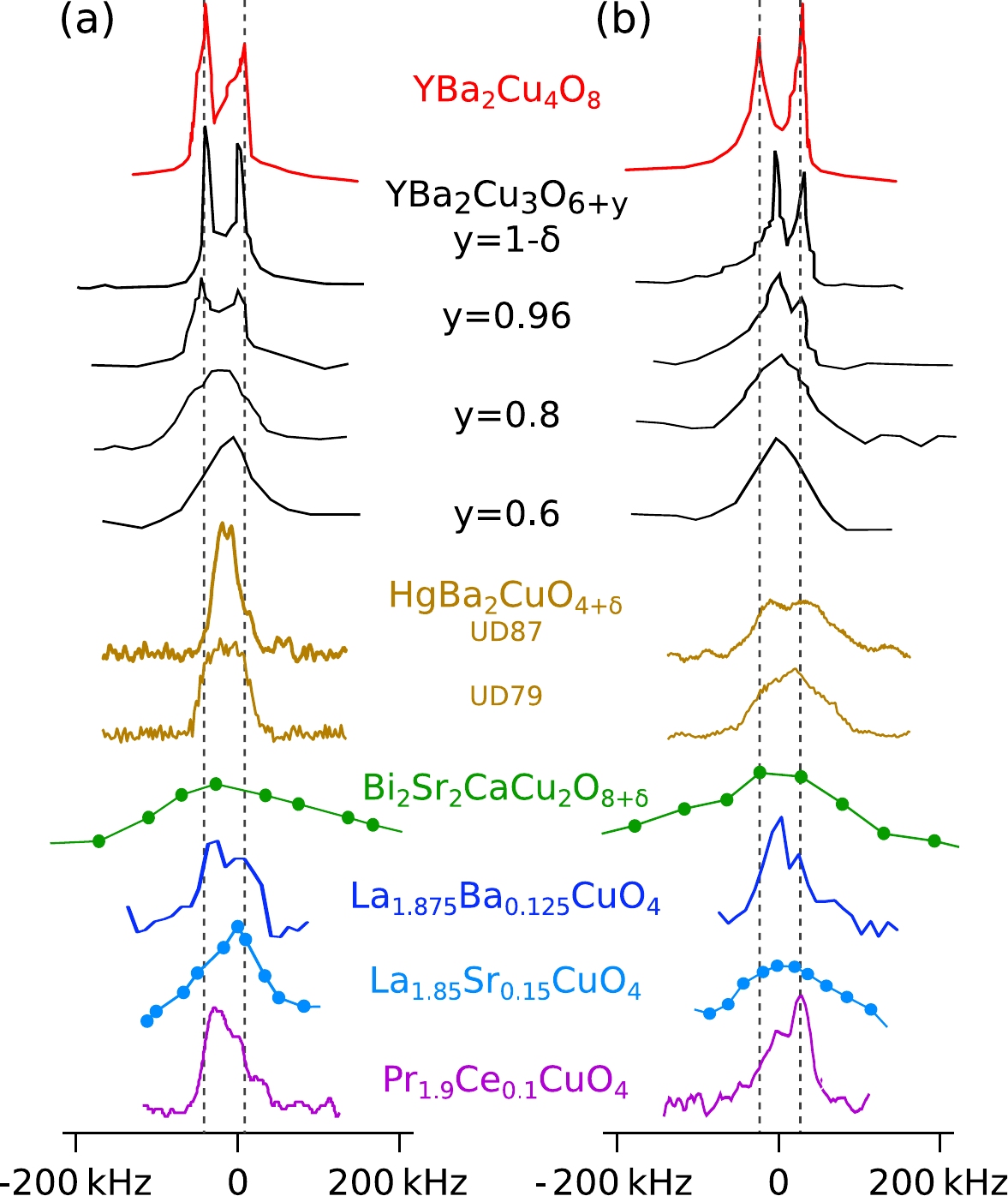}
\caption{Literature data of $^{17}$O NMR in various cuprates \cite{Crocker2011,Grafe2010,Haase2000,Lee2017,Suter2000,Takigawa1989,Yoshinari1990,Jurkutat2014}; displayed are the two outermost satellites for \cpara{}. The vertical dashed lines are defined by the splitting for \ybcoE. The similarities of the splittings and shapes suggest a common origin.}
 \label{fig:Fig8}
\end{figure}

Given similar NMR features in other YBCO materials, and preliminary experiments on \ybcoE{} and \ybcoF{} that also show increasing linewidths with pressure \cite{Jurkutat2017}, this charge ordering appears to be the missing link that explains why only a few cuprates have narrow NMR lines, and why linewidths can depend on the method of preparation, or imperfections \cite{Williams2001, Williams2007}. In particular, since the planar Cu NMR satellite linewidths are also known to be extremely sensitive to sample quality and doping level, e.g., for \cpara{} one finds values ranging from 0.1 to more than \SI{6}{MHz} \cite{Reichardt2016}, one is inclined to believe that the observed charge density variations can also affect the Cu $3d_{x^2-y^2}$ orbital. Given the larger pre-factor that relates the Cu linewidth to the related hole content \cite{Haase2004}, e.g., a 1\% variation of the Cu charge causes about \SI{1}{MHz} of linewidth for the Cu satellite (note that a population of the symmetric 4$s$ orbital does not affect the quadrupole splitting). Indeed, for an electron doped system it was shown that the Cu linewidths is given predominantly by a charge density variation in the $3d_{x^2-y^2}$ orbital \cite{Jurkutat2013}. Furthermore, similar linewidths and line splittings have been observed in a number of cuprates \cite{Grafe2010,Lee2017,Suter2000,Yoshinari1990,Reichardt2016,Jurkutat2014,Takigawa1989}, pointing to similar amplitudes of charge density variations. We collected literature $^{17}$O NMR spectra for various cuprates and plot them in Fig.~\ref{fig:Fig8}. These spectra clearly suggest that the charge ordering proven here with the two particular $^{17}$O NMR peaks could be ubiquitous (if some additional inhomogeneity is present, since NMR is a bulk local probe and takes the average over all nuclear positions).

Our results may raise many questions that cannot be answered right away. Not enough is known about this new charge order, its behavior at much higher or much lower temperatures, or as a function of doping, and its importance for cuprate physics in general. For instance, our sample is close to optimal doping and \tc{} increases by about \SI{4}{K} with pressure while the O charge ordering is enhanced. In addition, no simple trend between the NMR linewidths and the maximum \tc{} is found \cite{Rybicki2016}. 

Recent NMR reports on detwinned, underdoped YBCO samples observe similarly split O lines, that, however, were attributed to sites near empty and full chains \citep{Wu2013}.
Their observed (additional) Cu and O line splittings in high magnetic fields and at low temperatures in underdoped YBCO \cite{Wu2011, Wu2013} may point to a perturbation that affects the existing charge ordering, as well. We also know that moderate magnetic fields are able to induce Cu quadrupolar linewidth in \ybcoF{} \cite{Reichardt2016}, indicating an influence of the magnetic field that might be unexpected on general grounds.

At \SI{18}{kbar}, \SI{100}{K} we find in our sample fully established simple bulk ordering. Here, one may ask, which mechanism is responsible for selecting a particular unit cell orientation with respect to the field, cf. Fig.~\ref{fig:Fig7}~(f)? Is it a hidden symmetry that combined with pressure favors a field direction, or is it just the magnetic field itself that chooses the orientation of the charge pattern? We tried to answer the question by rotating the magnetic field, however, with limited experimental choices we could not produce a clear result and further experiments have to be performed. However, it is conceivable that if the magnetic field influences the charge distribution, its effect in the three different directions could be markedly different and influence our conclusions.

Finally, one would like to know about connections of our results to those obtained by other methods. We can only mention some ideas, as we are no experts for most of the large body of the literature. 
For instance, our charge density variation may correspond to the charge density wave (CDW) peaks observed in various x-ray studies of YBCO \cite{Ghiringhelli2012,Huecker2014,Comin2015a,Comin2016}. 
While the x-ray CDW peak is most pronounced in underdoped YBCO, at low temperature and when superconductivity is suppressed by a high magnetic field the feature is still found almost up to optimal doping, and it only smears out towards higher temperatures \cite{Blanco-Canosa2014}.
This was assigned to a decrease in correlation length below the detection limit where it might still be found with a local probe like NMR.
Furthermore, the observed local charge symmetry is similar to the electronic structure symmetries that were determined with x-ray \citep{Comin2015,Comin2015a}, but also with scanning tunneling microscopy (STM) \citep{Fujita2014, Mesaros2016}.

Our observation may also be related to early reports on inhomogeneities in the cuprates. For example, in \ybcoF{} two apical oxygen positions with similar probablility have been reported \cite{Leon1992}, or a CDW in La$_2$CuO$_{4.1}$ \cite{Lanzara1997}. Of course there are various other experimental findings and theoretical models that might relate to our results. Stripes and nematicity \cite{Zaanen1989,Tranquada1995,Kivelson2002} come to mind, or perhaps even orbital currents \cite{Varma2006} if they can lead to intra unit cell charge variations of the found size.

Very recently, some of us showed that the hitherto adopted interpretation of the NMR shift data in the cuprates was premature \cite{Haase2017}. We now know that, e.g., a large isotropic shift component is present in the cuprates on the strongly doped Fermi liquid side of the phase diagram. By lowering doping or temperature the isotropic component is lost, and the shifts become dominated by other contributions, which is not understood, currently. Therefore, it will be of interest to see how charge ordering discussed here relates to the cuprate shifts since we know already that at least some shift components are tied to the local charge density. In addition, the temperature dependence of the charge ordering observed here hints at motional effects that could easily cause quadrupolar relaxation, which was reported years ago \ybcoE{} \cite{Suter2000} and is of importance for the understanding of nuclear relaxation in these systems.

\section{Methods}

\textbf{Sample}
A high quality single crystal of \ybcoU{6.9} was used for high pressure $^{63}$Cu and $^{17}$O NMR experiments. It was obtained in the following way. 
Single crystals of \ybco{} were grown in non-reactive $\ce{BaZrO3}$ crucibles and were annealed afterwards at $\SI{300}{\celsius}$ in \SI{100}{bar} of O$_2$ for one week \cite{Erb1996}. This resulted in fully oxygenated single crystals ($y=1$) which are twinned within the $ab$-plane.
Some samples were cut into pieces of $\sim$150$\times$100$\times$100$\si{\micro m}^3$ for high pressure experiments. 
$^{17}$O exchange was performed at $\SI{600}{\celsius}$ in $90\%$ enriched $^{17}$O$_2$ gas for 3 days followed by a slow, stepwise cooling period of 1 month to $\SI{380}{\celsius}$.
The actual oxygen content was estimated \cite{Conder2001, Neumeier1993} from the measured superconducting transition temperature \tc(p=$0\,$kbar) = $90.3\,$K and its positive pressure gradient, i.e., \tc (p=$18\,$kbar) = $94.3\,$K, as well as comparison with ambient pressure NMR literature data.

\textbf{Crystal twinning}
The used single crystals are twinned, i.e., while the crystal $c$-axis is well defined, the $a$-axis and $b$-axis alternate over the whole crystal with a period of 1 to 3 $\mu$m. Twinning lines occur as angle bisectors due to tension at the boundaries, cf. inset of Suppl. Fig. 8. Those lines can easily be seen using polarization filter, cf. Fig. 8, and reveal that the alternating phases are homogeneously distributed for the used single crystals. In the investigated temperature and pressure regime no structural changes occur that can explain our observations. In particular, we can exclude detwinning as origin of the observed spectral changes at elevated pressure for various reasons. Detwinning is only achieved using uniaxial in-plane pressure at much higher temperature ($T > \SI{500}{K}$). Since we work with hydrostatic pressures and at much lower temperatures the chemical rearrangement of the Cu-O chains (O(1) diffusion) is not possible. In addition, de-twinned single crystals of YBa$_2$Cu$_3$O$_{6+y}$ are known to be stable over years and even at 600 K only partial re-twinning is observed, while we presented spectral changes for O and Cu NMR that are reversible in pressure and temperature and occur instantaneously. Furthermore, the spectral changes are inconsistent with a detwinning scenario. For instance, if one were to explain the data by orthorhombicity the axially symmetric EFG observed for $^{63}$Cu(2) at \SI{9}{kbar} and \SI{100}{K}, and at \SI{18}{kbar} and \SI{300}{K}, cf. Fig. 2 (e) and (c), would imply a mysterious loss of orthorhombicity altoghether, which is then re-established at  \SI{18}{kbar}, \SI{100}{K}, cf. Fig. 2 (f), while the persistent $^{17}$O double peak feature observed in all measurements for $c \parallel B_0$, cf. Fig. 4 (a), implies persistant orthorhombicity. Additional spectroscopic evidence against detwinning or other structural scenarios is provided by apical $^{17}$O(4) and chain $^{63/65}$Cu NMR data, cf. Suppl. Fig. 10. 


\textbf{NMR pressure cell}
Two B\"ohler-type moissanite anvils squeeze the sample space within a hardened nonmagnetic beryllium copper (BeCu) sheet that serves as a gasket. 
For an accurate alignment process and stable conditions under pressure the crystal is glued with pre-orientation on one of the anvils.
The NMR micro-coil was wrapped around the sample and paraffin oil was used as pressure medium.
We used the luminescence shift of ruby for pressure calibration. 
For details see Ref. \cite{Haase2009, Meier2015}.
The pressure cell was then fitted into a two-angle goniometer that was fastened on a home-build NMR probe.

\textbf{NMR and NQR experiments}
NMR and NQR experiments were carried out using standard pulsed spectrometers.
NQR experiments were done at zero field and the NMR measurement were performed in a superconducting magnet at \SI{11.7}{T}.
Spin echo pulse sequences were used to obtain spectra with frequency stepped experiments for broad lines and/or limited excitation/detection bandwidth.

\textbf{Alignment of the crystal}
Standard alignment procedures for single crystals within the pressure cell are not possible due to sample size and limited visible access.
A two angle goniometer allows the alignment of the sample with the magnetic field $B_0$ perpendicular to the CuO$_2$ plane (\cpara) and  parallel to half of the planar Cu-O-Cu bonds, i.e., parallel to both $a$ and $b$ axis in a twinned crystal (\abpara).
For this, we used the large angular dependence of the magnetic shift and quadrupole frequency of planar Cu and O whose principle axis coincide with the crystal axis.
The narrow planar $^{63}$Cu central transition ($35-70\,$kHz) changes under rotation by $\approx 3\, $MHz and exhibits two maxima for \cpara{} and \cperp{}.
Starting from \cperp{} the planar $^{17}$O satellite frequencies reveal extrema for $B_0$ parallel to either planar Cu-O-Cu bond direction (\abpara).
See Suppl.~Fig.~9 for details of the alignment procedure.

\textbf{Electric field gradients, NQR and NMR}
An electric field gradient (EFG) tensor at a particular nuclear site is characterized by its principle values $V_{XX}$, $V_{YY}$ and $V_{ZZ}$ and its orientation. $V_{ZZ}$ is always the largest component (the sign cannot be determined), and since the tensor is traceless, one typically defines $|V_{ZZ}|\geq |V_{YY}| \geq |V_{XX}|$. However, for Cu in the cuprates Z is clearly along the crystal $c$ axis, due to the nearly vanishing asymmetry the other two axes could switch from place to place if one assumes the above convention. Therefore, we abstain from using the convention and rather speak about the spectral appearance of lines ($\Delta f_\pm$) for Cu NMR and \cperp{}.

The interaction of the EFG with the nuclear electric quadrupole moment $eQ$ gives rise to an energy splitting that can be measured even in the absence of an external magnetic field with NQR (nuclear quadrupole resonance). For planar oxygen the splitting is too small ($\approx 1\,$MHz) and NQR measurements are not feasible with the small crystallites used for the pressure cell. However, planar $^{63}$Cu exhibits a rather large quadrupole interaction ($\approx 31\,$MHz) and one obtains a single line at $\nu_{\mathrm{Q}}(\mathrm{NQR})=\kappa V_{ZZ}(1+\eta^2/3)^{1/2}$ with  $\kappa=3eQ/2I(2I-1)h$ and $I$ the nuclear spin. 
Since the asymmetry parameter $\eta$ is small ($\eta<0.02$), the line position and it's width measure the largest principle value $V_{ZZ}$ and its spatial distribution $\Delta V_{ZZ}$, respectively.

For NMR in typical magnetic fields (here \SI{11.7}{T}), the quadrupole interaction perturbs the Zeeman interaction, which results in $2I$ resonance lines, i.e., there is a central transition and $2I-1$ satellite transitions, cf. Fig.~\ref{fig:Fig1}~(b).
The principle components of the EFG ($ V_{ii}$) are measured by aligning the magnetic field along the principle axes ($i=X,Y,Z$), cf. Suppl.~Fig.~2.
Besides higher order effects (up to third order were taken into account \cite{Wolf1970} analytically), the splittings along the principle axes are given by $\nu_{\mathrm{Q}}(\mathrm{NMR})=\kappa V_{ii}$.

For planar Cu the magnetic contribution to the satellite linewidths is negligibly small so they measure the spatial distribution of $V_{ii}$.
However, for planar O, magnetic and quadrupole effects are of similar size and are not easily separable for the observed double peak satellites (without a particular model). 
Thus, the comparison of the outer most satellites with the central transition (having vanishing quadrupole contributions) estimates the spatial EFG variation.\\
\textit{Planar Copper:} From the local symmetry of planar Cu one knows that the EFG is nearly axially symmetric, i.e., $V_{XX} \approx V_{YY} \approx -V_{ZZ}/2$. Then, for \abpara{} the deviation from $V_{ZZ}/2$ of the measured spectra gives the actual EFG asymmetry, i.e., spectral weight below $|V_{ZZ}|/2$ correspond to nuclei with $X\parallel B_0$ ($|V_{XX}|<|V_{ZZ}|/2$) and above to nuclei with $Y\parallel B_0$ ($|V_{YY}|>|V_{ZZ}|/2$).
As mentione above, since Cu nuclei are observed with both $V_{YY}$ and $V_{XX}$ in magnetic field direction, we use for a simple notation $\Delta f_{\pm} =\pm \epsilon$ with $\epsilon=\kappa \eta V_{ZZ} =\kappa (V_{XX}-V_{YY}$) for the deviation in frequency units from the axially symmetric case, cf. Suppl.~Fig.~2. 
Note, that the EFG asymmetry is described by the parameter $\eta=(V_{XX}-V_{YY})/V_{ZZ}$ that is usually defined to be positive.
Any spatial variation $\Delta V_{ZZ}$ gives rise to satellite linewidth of $\kappa \Delta V_{ZZ}/2$ and again, any deviation would reflect additional spatial variations of $\Delta V_{XX}$ or $\Delta V_{YY}$.\\
\textit{Planar Oxygen:}  Since the crystal lacks rotational symmetry about the Cu-O-Cu bond, the EFG asymmetry at planar O is substantial and we find different principle values along the Cu-O-Cu bond, perpendicular to it and along the $c$-axis. 
Thus, for \abpara{} we measure clearly separated half of the planar O along the bond ($\sigma \parallel B_0$) and half perpendicular ($\sigma \perp B_0$) to it, cf. Suppl.~Fig.~1 and Suppl.~Fig.~2.
However, the oxygen positions are in principle not identical (measured EFG components do not belong to the same oxygen position).
Thus, from statistical quantities of the outer most satellite lines (quadrupole contributions is largest and therefore this transition is most reliable), e.g., center of gravity, edges and peaks, measured perpendicular to bond and for \cpara, one can use Laplace equation to calculate the expected quantities along the bond, cf. Suppl.~Fig.~2. 
Disagreement is a simple test for equivalence of oxygen sites.

\textbf{Simulation of NMR spectra}
Since NQR on planar Cu measures directly the variation of the largest principle component ($\eta \ll1$), we simulate the planar Cu NMR spectra based on this measured distribution by diagonalization of the Hamiltonian.
For this, we estimate the EFG asymmetry from the pronounced satellite peak observed for \abpara{}; the magnetic shift was adjusted such that all observed transitions coincide with the simulation.
We compare the observed NMR satellite pattern observed for \cpara{} and \abpara{} with the simulated one assuming two EFGs switched by 90\degree{} as expected for a twinned sample, cf. Suppl.~Fig.~3 and Fig.~4. 
The intensities were adjusted such that the peak values match and crystal twinning results in equal intensities.
We neglect magnetic shift variations.

\textbf{Quantitative determination of local Cu and O charges} is provided by their NMR quadrupole splittings that are determined by the hole contents in the bonding orbitals Cu $3d_{x^2-y^2}$ ($n_d$) and O $2p_\sigma$ ($n_p$) \cite{Haase2004,Jurkutat2014}:
\begin{widetext}
\begin{align}
^{63}\nu_{\rm{Q},c}&=\underbrace{94.3{\rm ~MHz}}_{\substack{^{63}\nu_{\rm Q,c} \text{ for}\\ \text{Cu 3d}^9}} \cdot n_d - \underbrace{14.2 {\rm ~MHz}}_{\substack{^{63}\nu_{\rm Q,c} \text{ for}\\ \text{1 e- in Cu 4p}}} \cdot \underbrace{\beta^2}_{\substack{\text{overlap}\\ \langle \text{Cu 4p} \vert \text{O 2p} \rangle}} \cdot \underbrace{4(2-n_p)}_{\substack{\text{charge in 4}\\ \text{bonding O 2p$_\sigma$}}} + \underbrace{^{63}C_c}_{\substack{\text{other contributions}\\ \text{}\approx 0}}\label{EHS63}\\
&\nonumber\\
^{17}\nu_{\rm{Q},\sigma}&=\underbrace{2.45{\rm ~MHz}}_{\substack{^{17}\nu_{\rm Q} \text{ for}\\ \text{O 2p}^5}} \cdot n_p +  \underbrace{^{17}C_\sigma}_{\substack{\text{other contributions}\\ = 0.39\text{~MHz}}}
\label{EHS17}
\end{align}
\end{widetext}

\begin{wrapfigure}{r}{0.2\textwidth}
  \vspace{-15pt}
\begin{center}
\includegraphics[scale=0.75]{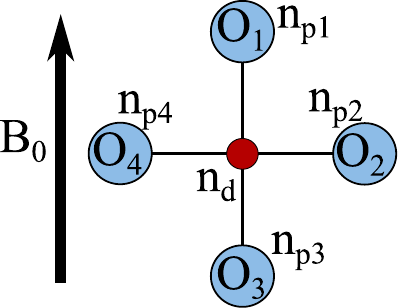}
\end{center}
  \vspace{-10pt}
\caption[Local CuO$_4$ unit]{Local CuO$_4$ unit consisting of one Cu with its hole content $n_\mathrm{d}$ and four surrounding O atoms with $n_{\mathrm{p},1...4}$.}
 \label{fig:Fig9}
\end{wrapfigure}

Equations \eqref{EHS63} \& \eqref{EHS17} describe the splitting of Cu and O for the respective principle axis of charge symmetry, i.e., $ c \parallel B_0$ for Cu and along the $\sigma$-bond for O. 
Since the asymmetry of the Cu EFG is changing with pressure this must be related to a change of the surrounding O charges, cf. Fig.~\ref{fig:Fig9}.
For instance, the charge $(2-n_{p1})$ of O$_1$ sigma bonding orbital along the in-plane field is seen by the Cu nucleus as partial occupation of the nominally empty Cu 4p along $\overline{\text{O}_1\text{-Cu-O}_3}$. 
One electron in 4p causes a Cu splitting along the orbital axis of symmetry of +28.4~MHz, and -14.2~MHz in perpendicular directions, i.e., $c \parallel B_0$ or the perpendicular in-plane direction.
So with the field in the plane along the bonding orbitals of O$_1$ and O$_3$, and with the overlap of Cu 4p and O 2p$_\sigma$ of $\beta^2=0.4$, the charge on the four surrounding O contributes to the Cu splitting as $+28.4$~MHz$\cdot \beta^2 \cdot (4-n_{p,1}-n_{p,3})-14.2$~MHz$\cdot \beta^2 \cdot (4-n_{p,2}-n_{p,4})$. 
If all charges correspond to the average O hole content, $n_{p,1...4}=\left\langle n_p \right\rangle$, the contribution is simply minus half that seen in the $c$-direction, i.e., $+7.1$~MHz$\cdot \beta^2 \cdot (8-n_{p})$, cf. Eq.~\eqref{EHS63}, and the O contribution to the Cu splitting shows c-axial symmetry.
If however, the hole contents differ, $n_{p,i}=\left\langle n_p \right\rangle + \delta_i$, and in an asymmetric fashion, i.e. $\delta_1+\delta_3 \neq \delta_2+\delta_4$, the Cu satellite is shifted from the position for an axially symmetric EFG by $-28.4 \cdot \beta^2 \cdot ( \delta_1+\delta_3-\frac{\delta_2+\delta_4}{2})$.
Note that as the Cu splitting per our definition is positive for $c \parallel B_0$, it is negative for \abpara{} such that the measured splitting is increased if $\delta_1+\delta_3 >2({\delta_2+\delta_4})$.
Also note that as long as we have charge balance $\Sigma \delta_i =0$ around all Cu, the O variation will not affect the Cu linewidth measured in $ c \parallel B_0$.

Using the O satellite splitting in $c \parallel B_0$ as a measure of the amplitude of the O hole content variation ($\pm \delta$), we find $2 \delta = {\Delta ^{17}\nu_{\mathrm{Q},c}}/{1.227 \mathrm{MHz}}$ \cite{Haase2004}. 
If we take, for instance, the O satellite line splitting seen at ambient conditions of $\Delta ^{17}\nu_{Q,c} ( \text{0~kbar, 300~K})= 18~\text{kHz}$, cf. Fig.~\ref{fig:Fig4}, we find $\delta=0.73\%$.
In the charge ordered configuration depicted in Fig.~\ref{fig:Fig7}~(b), i.e., $\delta_1=\delta_3=-\delta_2=-\delta_4=\delta$, this value of $\delta$ yields an increase in the measured Cu splitting ($B_0 \parallel$ O$_1$-Cu-O$_3$-bond) of +250~kHz.
If the local charge order was in the other direction, i.e.,  $-\delta_1=-\delta_3=\delta_2=\delta_4=\delta_0$, the effect would be a decrease of the measured Cu splitting by -250~kHz.

\section{Acknowledgements}
We thank B. Fine, R. Reznicek and C. Mazzoli for helpful discussions of our results, T. L\"uhmann and J. Barzola-Quiquia for microscopy imaging. 
We acknowledge the financial support from the University of Leipzig, the free state of Saxony, the European Social Fund (ESF), and the Deutsche Forschungsgemeinschaft (DFG).


\bibliography{JHPRX}
\newpage
\includepdf[pages={1}]{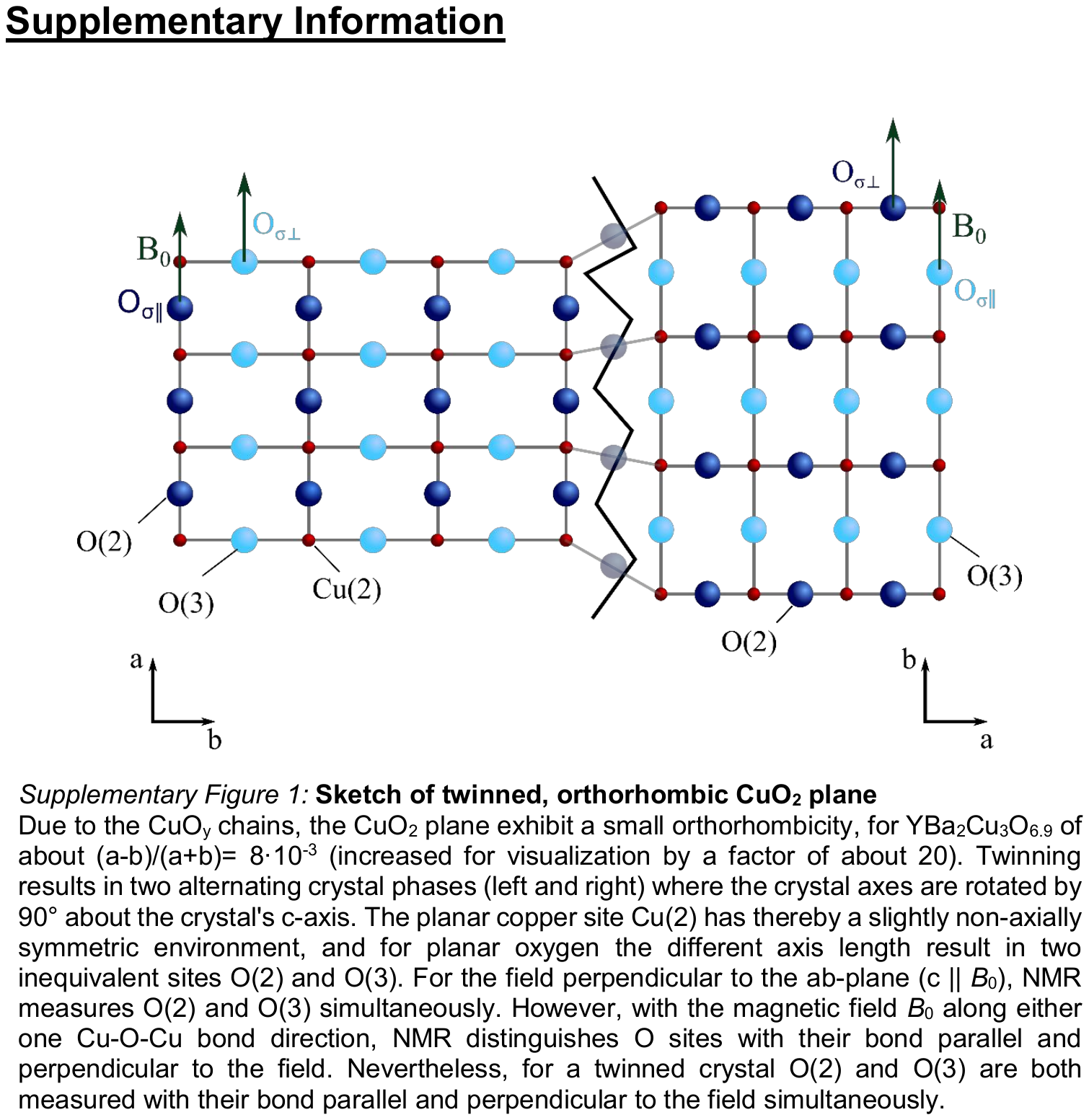}
\hspace{0.2cm}
\newpage
\includepdf[pages=2]{Supplement.pdf}
\hspace{0.2cm}
\newpage
\includepdf[pages=3]{Supplement.pdf}
\hspace{0.2cm}
\newpage
\includepdf[pages=4]{Supplement.pdf}
\hspace{0.2cm}
\newpage
\includepdf[pages=5]{Supplement.pdf}
\hspace{0.2cm}
\newpage
\includepdf[pages=6]{Supplement.pdf}
\hspace{0.2cm}
\newpage


\section*{Supplementary material}

# Supplementary Information

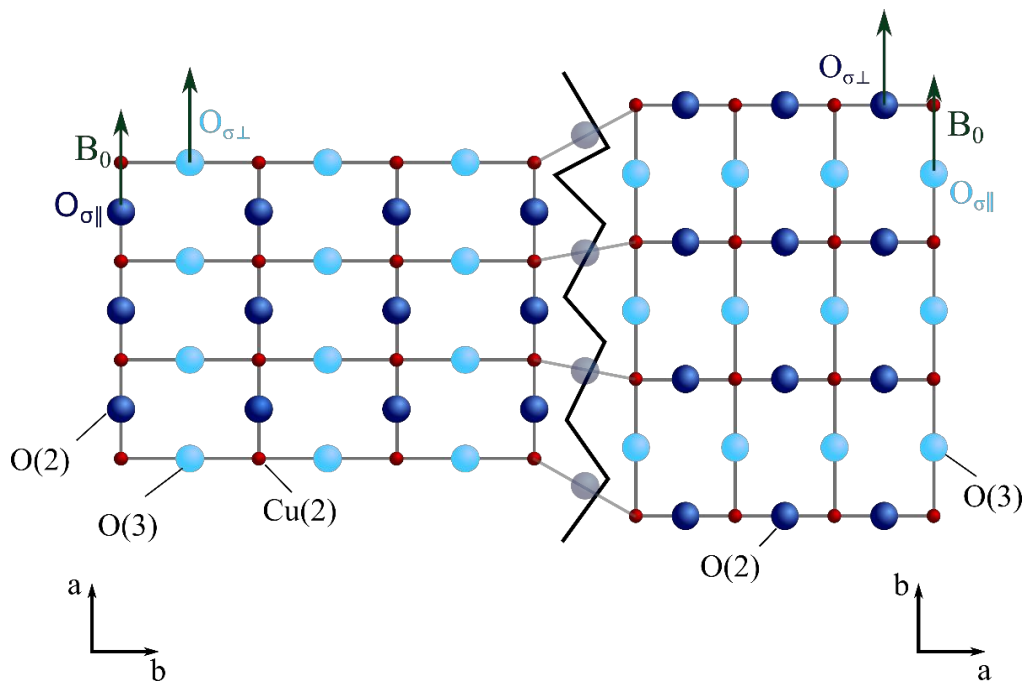

*Supplementary Figure 1:* **Sketch of twinned, orthorhombic CuO$_2$ plane**
Due to the CuO$_y$ chains, the CuO$_2$ plane exhibit a small orthorhombicity, for YBa$_2$Cu$_3$O$_{6.9}$ of about (a-b)/(a+b)= 8·10$^{-3}$ (increased for visualization by a factor of about 20). Twinning results in two alternating crystal phases (left and right) where the crystal axes are rotated by 90° about the crystal's c-axis. The planar copper site Cu(2) has thereby a slightly non-axially symmetric environment, and for planar oxygen the different axis length result in two inequivalent sites O(2) and O(3). For the field perpendicular to the ab-plane (c || $B_0$), NMR measures O(2) and O(3) simultaneously. However, with the magnetic field $B_0$ along either one Cu-O-Cu bond direction, NMR distinguishes O sites with their bond parallel and perpendicular to the field. Nevertheless, for a twinned crystal O(2) and O(3) are both measured with their bond parallel and perpendicular to the field simultaneously.

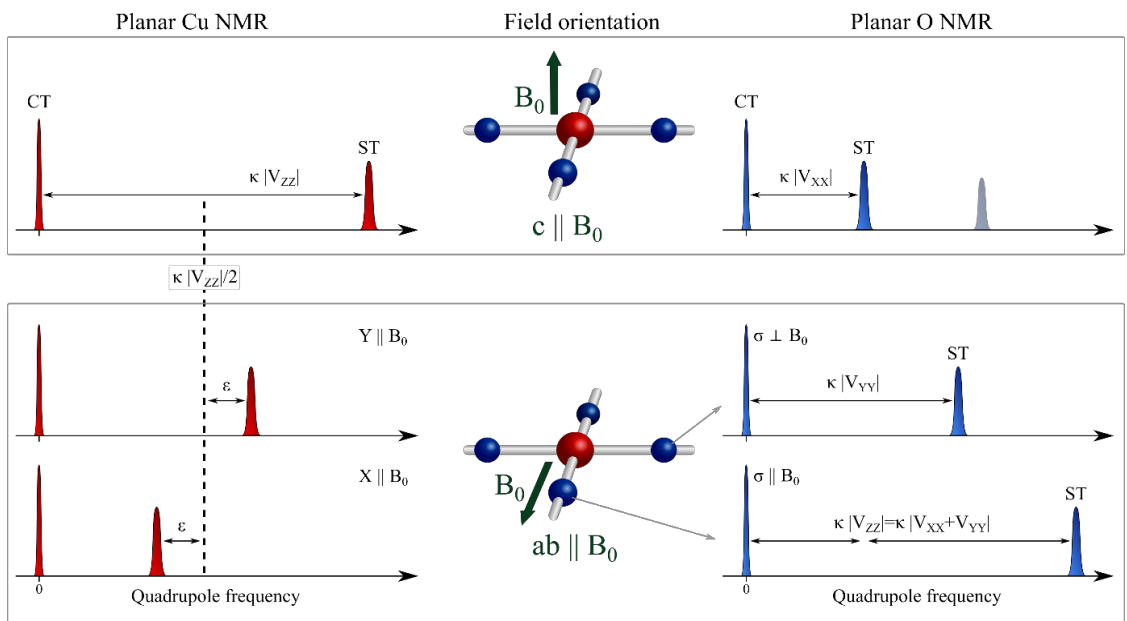

*Supplementary Figure 2:* **Electric field gradient at planar Cu and O measured by NMR**
The principle components of the planar Cu and O EFGs are measurable as frequency shifts between the central transition (CT) and the satellite transitions (ST). (left) The largest EFG principle value $V_{ZZ}$ of planar Cu is measured for c || $B_0$ ($\kappa = 3eQ/2I(2I-1)h$ ). For ab || $B_0$, the satellite position of an axially symmetric EFG ($V_{XX}=V_{YY}$) is indicated by the dashed line. A small non-axial symmetry of the EFG ($V_{XX} \neq V_{YY}$) shifts the resonance from the dashed line towards higher (lower) quadrupole frequencies by $\varepsilon = \kappa |V_{XX}-V_{YY}|$ if the magnetic field is along the EFG principle axis Y (X). If both peaks are measured simultaneously, the orientation of the EFGs vary in field direction. (center) Respective magnetic field orientation illustrated at a planar CuO$_4$ unit. (right) The smallest O EFG principle value $V_{XX}$ is measured for c || $B_0$. With the magnetic field along either one Cu-O bond direction (ab || $B_0$), $V_{YY}$ is measured for the O having their bond perpendicular to magnetic field ($\sigma \perp B_0$) and the largest component $V_{ZZ}$ is obtain for the O with the field parallel to the bond ($\sigma$ || $B_0$). Due to the traceless of the EFG, $V_{ZZ}$ can also be calculated from the two other directions.

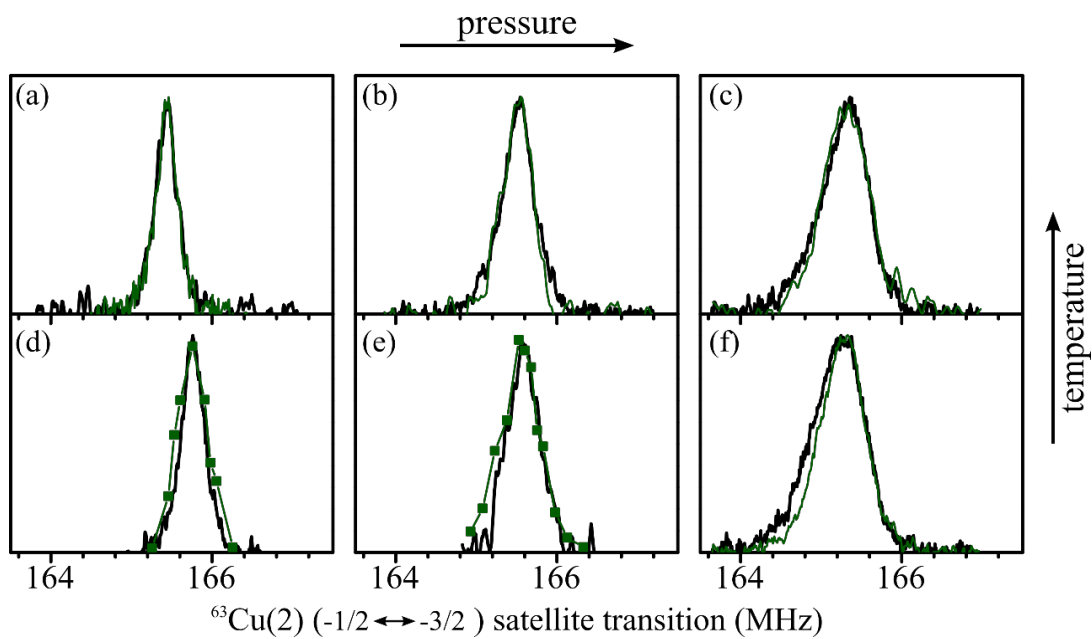

*Supplementary Figure 3:* **Planar $^{63}$Cu NQR vs. satellite spectra for c || $B_0$**
The upper and lower panels correspond to 300 K and 100 K, respectively, with pressures of 0 kbar (a, d), 9 kbar (b, e) and 18 kbar (c, f). The NMR (c || $B_0$) satellite transitions (black) agree with simulated spectra (green) based on NQR ($B_0$ = 0) measurements, cf. Methods in main article. This shows that the largest principle value of the Cu EFG is well-defined in terms of mean value and distribution.

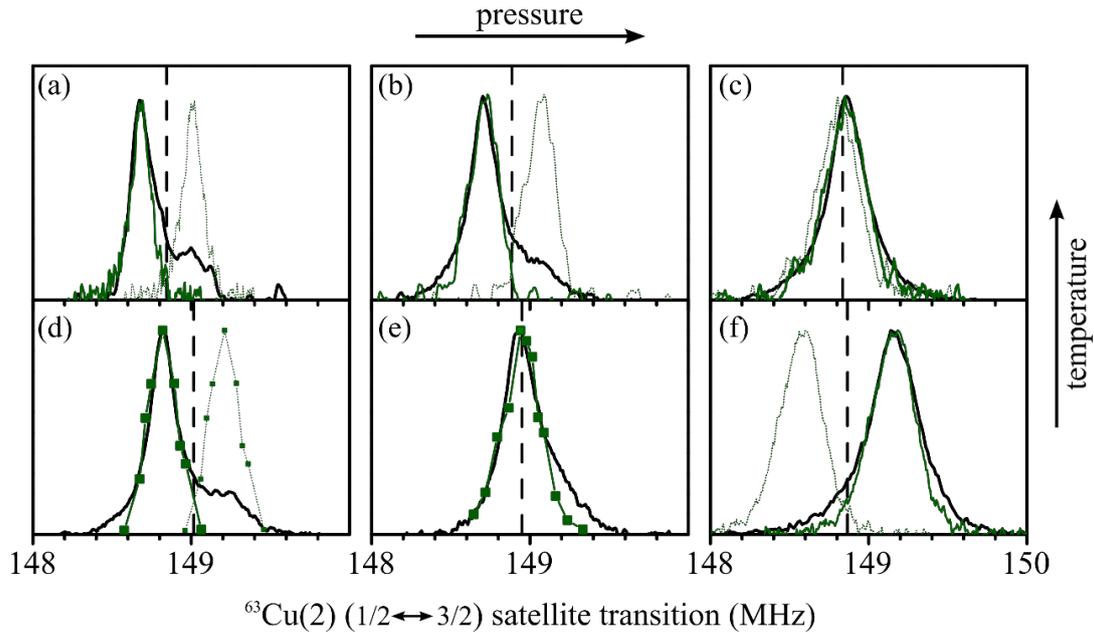

*Supplementary Figure 4:* **Planar $^{63}$Cu NQR vs. NMR satellite spectra for the magnetic field $B_0 \perp c$ along one Cu-O-Cu bond direction**
The upper and lower panels correspond to 300 K and 100 K, respectively, with pressures of 0 kbar (a, d), 9 kbar (b, e) and 18 kbar (c, f). Vertical dashed lines indicate the calculated position expected for an axially symmetric EFG from zero field data. The NMR satellites (black) do not agree with what is expected from NQR measurements and an EFG asymmetry expected for orthorhombicity in a twinned crystal (green), cf. Methods in main article. For instance, the intensity mismatch (a, b, d) or the entire absence of an expected second line (f) is direct evidence again the orthorhombic scenario for the asymmetry of the planar Cu EFG. In particular the coincidence in (f) of the NMR satellite with only one, the upper green line shows alignment of Cu EFG throughout the twinned crystal, i.e., charge ordering.

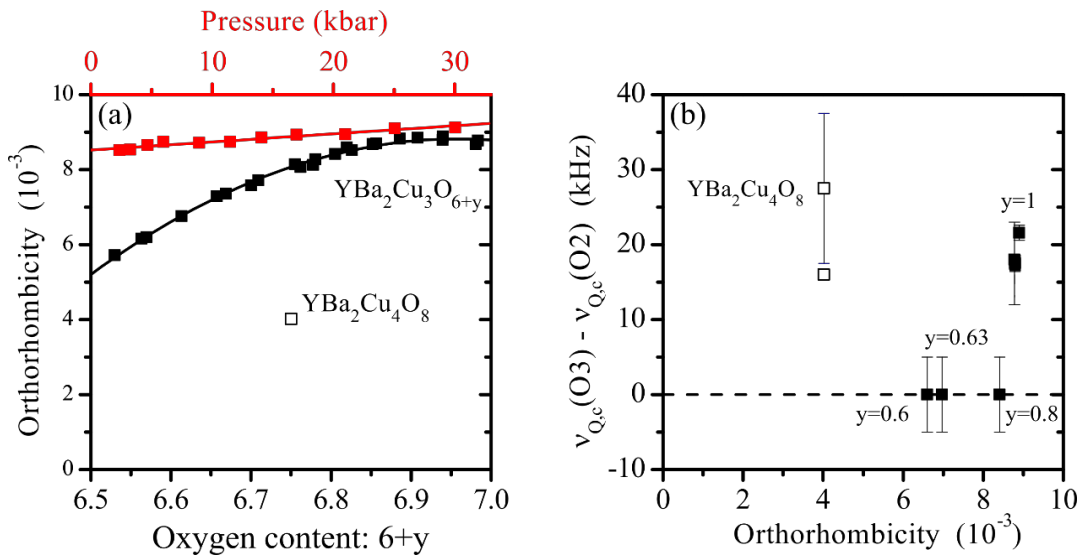

*Supplementary Figure 5:* **Orthorhombicity and planar oxygen line splitting**
(a) Monotonic increase of the orthorhombicity (a-b)/(a+b) with doping (black). Also shown is the underdoped (stoichiometric) component $YBa_2Cu_4O_8$. Pressure (red) changes the orthorhombicity of optimally doped YBCO only slightly. (b) Difference of the planar oxygen quadrupole splittings of the two inequivalent sites O(2) and O(3) measured in c || $B_0$ vs. orthorhombicity. Satellite splitting seems to occur only in well-ordered compounds and, although the orthorhombicity differs by a factor of two, the size of the splitting is similar in these materials. Interestingly, the total satellite linewidth is largely independent on whether a splitting is observed or not, cf. Fig. 8 in main article.

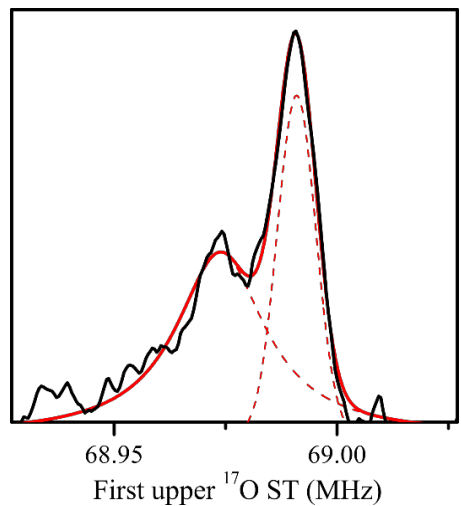

*Supplementary Figure 6:* **$^{17}$O satellite lineshape**
The planar $^{17}$O first high frequency satellite transition (black) at ambient pressure measured with the magnetic field along the Cu-O bond direction reveals only one sharp edge to higher quadrupole frequencies. Also shown is a fit of a distribution function (solid red) consisting of a Gaussian and Lorentzian line (dashed red) with roughly equal intensities. If solely two lines are present, they have different shapes, but also a more complex distribution could be present.

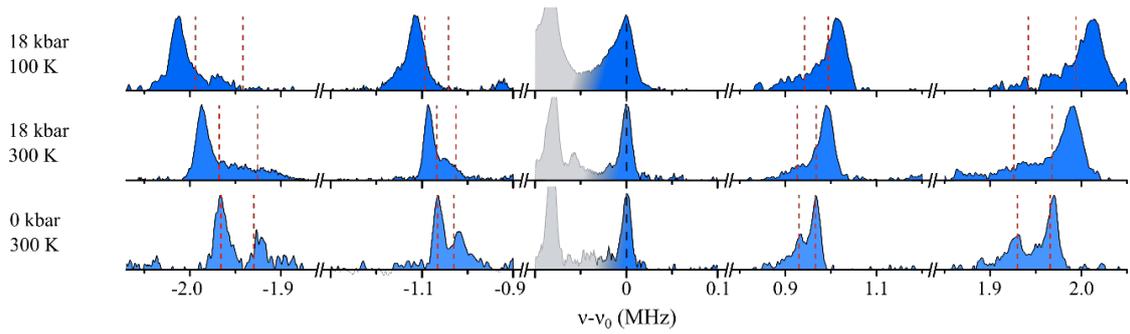

*Supplementary Figure 7:* **Pressure and temperature dependence of planar $^{17}$O spectra with the magnetic field $B_0 \perp c$ along one Cu-O-Cu bond direction**
The magnetic frequency contribution was subtracted to emphasize the changes in the quadrupole splitting. Red dashed lines are the expectations using Laplace equation from the peak values of quadrupole splitting measured in the other two orientations, i.e., with the magnetic field perpendicular to the Cu-O-Cu bond and parallel to the c-axis. While we have agreement at ambient conditions, the pronounced differences at elevated pressure reflect local charge ordering. Transitions from other O sites are shaded.

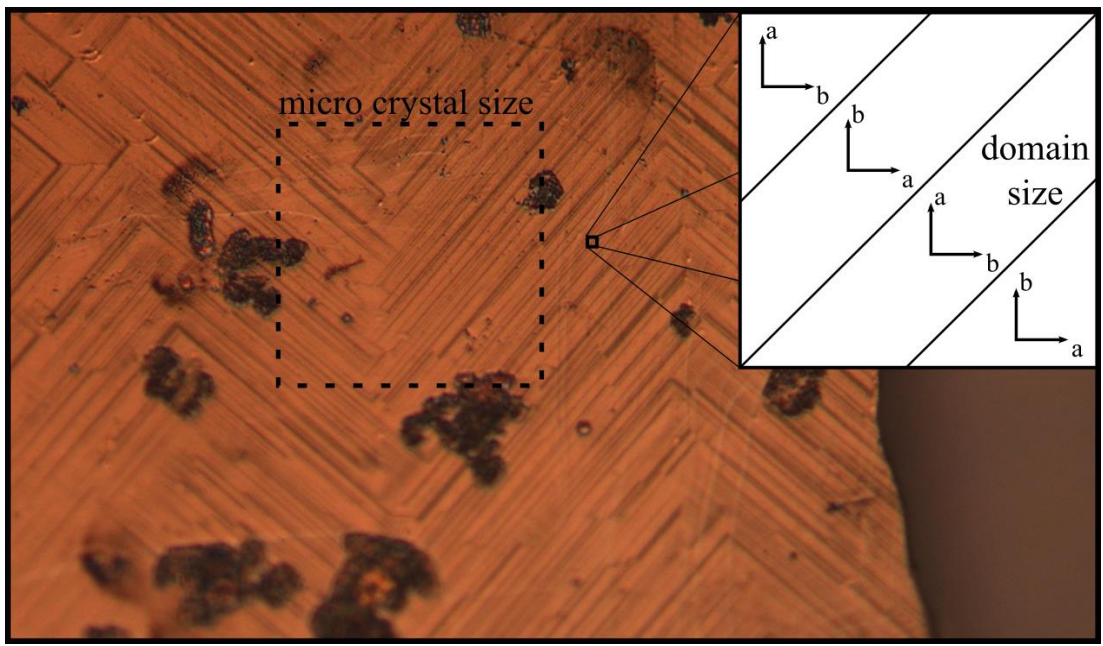

*Supplementary Figure 8:* **Polarization microscopy image of a YBa$_2$Cu$_3$O$_7$ single crystal**
(*Main panel*) Twinning lines, which occur at the twin boundaries on the crystal's ab-plane, are visually enhanced using a polarization filter. A square of 150 µm illustrates the micro sample dimensions and numerous boundaries indicating equal occurrence of both orientations. (*Inset*) Pictogram of the visible twinning lines between domains (1-3 µm) and the corresponding orientation of the alternating crystal axes.

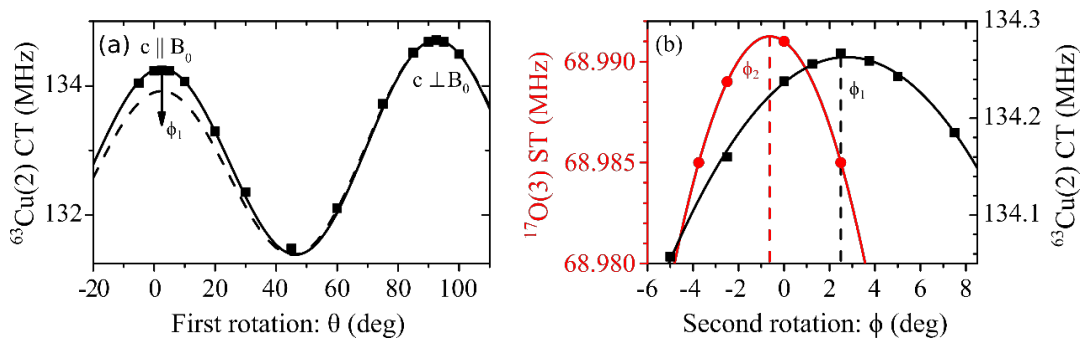

*Supplementary Figure 9:* **Alignment of the crystal within the pressure cell using NMR**
The single crystal is clued on the anvil and the assembled pressure cell is mounted on a two angle goniometer. The alignment of the crystal with respect to the external magnetic field was done using the angular dependence of the magnetic shift and quadrupole frequency. (a) $^{63}$Cu central transition as a function of the first rotation angle $\theta$ exhibits two maxima. While one maximum corresponds to $c \perp B_0$, only if the rotation axis is perpendicular to the crystal's c-axis (solid line), the other maximum gives $c \parallel B_0$. A misalignment of $\Phi_1 = 10°$ is shown by the dashed line demonstrating the good pre-orientation of the crystal. (b) Fine alignment using a second rotation axis which is perpendicular to the external magnetic field direction and the first rotation axis. For $c \parallel B_0$, another maximum appears for $^{63}$Cu central transition at $\Phi_1$ (black). Starting at $c \perp B_0$, the maximal frequency of the satellite transition of $^{17}$O at $\Phi_2$ is used to find $B_0$ along one Cu-O-Cu bond direction. The solid lines in (b) are parabolic fits.

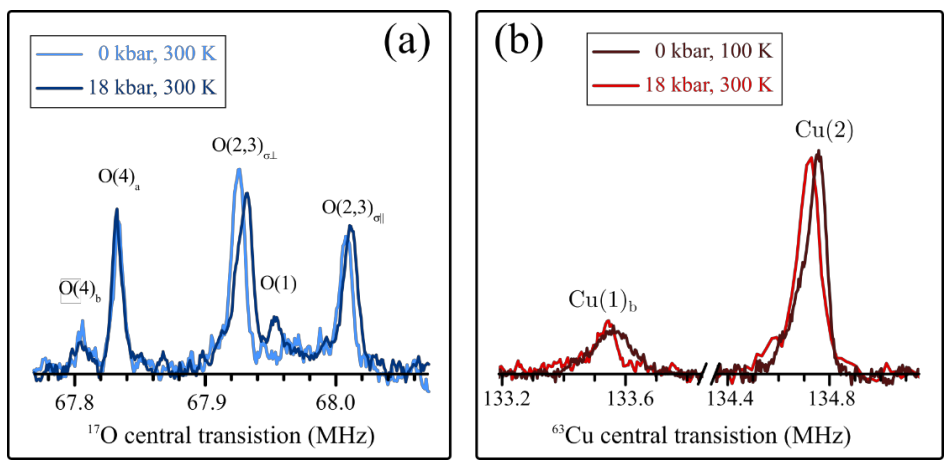

*Supplementary Figure 10:* **Unchanged twinning seen by apical O and chain Cu**
(a) $^{17}$O central transitions in $c \perp B_0$ without and with applied pressure. Signal intensity of apex oxygen measured along a and b axis unchanged by application of 18 kbar, showing that the crystal's twinning is unchanged. Signal intensity of O(4)$_b$ is suppressed by long spin-lattice relaxation time relative to short experimental repetition time.
(b) $^{63}$Cu central transitions in $c \perp B_0$ of chain copper Cu(1)$_b$, measured along CuO-chain, and planar Cu(2) without and with applied pressure. The integrated spectral intensity ratio $I_{Cu(2)}/I_{Cu(1)b}$ is 3.5 at (0 kbar, 100 K) and 4.4 at (18 kbar, 300 K), consistent with the expected ratio of 4 for a twinned crystal rather than 2 for a de-twinned crystal. At (18 kbar, 300 K) suspected overlap of the central transition of Cu(1)$_a$ with Cu(2) could explain the increased ratio since then the expected ratio would be $(I_{Cu(2)}+I_{Cu(1)a})/I_{Cu(1)b}=5$.



\end{document}